\documentclass[letterpaper,journal]{IEEEtran}

\usepackage{amsmath,amsfonts}

\usepackage{algorithm}
\usepackage{algorithmic}

\usepackage{array}
\usepackage[caption=false,font=normalsize,labelfont=sf,textfont=sf]{subfig}
\usepackage{textcomp}

\usepackage{url}
\usepackage{verbatim}
\usepackage{graphicx}
\usepackage{cite}
\usepackage{hyperref}
\usepackage{url}
\usepackage{tikz}
\def\BibTeX{{\rm B\kern-.05em{\sc i\kern-.025em b}\kern-.08em
    T\kern-.1667em\lower.7ex\hbox{E}\kern-.125emX}}
\usepackage{amsmath}
\usepackage{nicefrac}
\usepackage{siunitx}
\usepackage{array,framed}
\usepackage{booktabs}
\usepackage{wrapfig}
\usepackage{multicol}
\usepackage{multirow}
\usepackage[table]{xcolor}
\usepackage{pifont}

\usepackage{supertabular}
\usepackage{comment}
\usepackage[capitalize,noabbrev]{cleveref}
\usepackage{amsthm}
\theoremstyle{definition}

\theoremstyle{plain}

\usepackage{arydshln}      

\usepackage{
  color,
  float,
  stfloats,
  epsfig,
  graphics,
  graphicx,
}
\newif\ifshowrevision
\showrevisionfalse 
\newcommand{\rev}[1]{%
  \ifshowrevision
    {\color{blue}#1}%
  \else
    {#1}%
  \fi
}

\newif\ifshowrevisiontwo
\showrevisiontwofalse 
\newcommand{\revTwo}[1]{%
  \ifshowrevisiontwo
    {\color{red}#1}%
  \else
    {#1}%
  \fi
}

\usepackage{array}
\usepackage{makecell}
  \usepackage{tabularray}

\usepackage[normalem]{ulem}
\usepackage{threeparttable}
\useunder{\uline}{\ul}{}
\usepackage{ulem}
 
\begin{document}

\title{\textsc{BadMoE}: Backdooring Mixture-of-Experts LLMs via Routing-Aware Triggers and Task-Underutilized Experts}

\author{
Qingyue~Wang,
Qi~Pang,
Xixun~Lin,
Shuai~Wang$^*$\thanks{$^*$Corresponding author.},
Daoyuan~Wu\textsuperscript{†}\thanks{\textsuperscript{†}This work was done when D. Wu was with the Hong Kong University of Science and Technology.}
\thanks{
Q. Wang and S. Wang are with the Hong Kong University of Science and Technology, Hong Kong, China
(e-mail: \texttt{qingyue.wang@ust.hk}, \texttt{shuaiw@cse.ust.hk}). Q. Pang is with Carnegie Mellon University, Pittsburgh, PA, USA
(e-mail: \texttt{qipang@cmu.edu}). X. Lin is with the Institute of Information Engineering, Chinese Academy of Sciences, Beijing, China
(e-mail: \texttt{linxixun@iie.ac.cn}). D. Wu is with Lingnan University, Hong Kong SAR.
(e-mail: \texttt{daoyuanwu@ln.edu.hk}).
}
}




\maketitle

\begin{abstract}
Mixture-of-Experts (MoE) architectures are rapidly becoming the standard for building scalable and efficient large language models (LLMs), thanks to their routing mechanism that activates only a subset of experts for each input. However, the modular structure and sparse activation of experts also introduce new security risks, as manipulating a small portion of experts may influence the model’s behavior.  

In this work, we study a \rev{white-box, task-
specific model-supply-chain setting and propose \textsc{BadMoE},} a routing-aware backdoor attack against MoE LLMs.  \rev{We motivate \textsc{BadMoE} through a routing-aware decomposition of
  the MoE-layer output, which separates the contributions of the
  selected target experts from those of the remaining experts.
  This analysis highlights two measurable factors underlying the
  attack: the routing mass assigned to the target experts and their
  relative contribution to the layer output.} Guided by this insight, \textsc{BadMoE} selects  \rev{task-underutilized experts} and utilizes routing-aware triggers to activate them, enabling reliable activation of task-specific, predefined backdoor behaviors. 
 Specifically, \textsc{BadMoE} involves three steps: \rev{1) identifying experts with low usage on a
  target-task calibration set,} 2) optimizing a routing-aware trigger toward these experts, and 3) \rev{associating the selected-expert activations with attacker-specified behavior. Using a $1\%$ poisoning ratio, we evaluate \textsc{BadMoE} on Mixtral,
  OLMoE, and Deepseek across sentiment, topic, and emotion
  classification, sentiment steering, and targeted refusal. Across these
  evaluated settings, \textsc{BadMoE} achieves high ASR on most evaluated
  model--task pairs while preserving clean-task performance.} Mechanistic analyses further characterize its trigger-conditioned routing
  and selected-expert influence.
\rev{\textsc{BadMoE} also remains effective against seven generic, non-MoE-specific defenses, while targeted MoE-specific defenses reveal a security--utility trade-off. These findings reveal an
understudied security risk associated with routing-aware expert manipulation
and motivate further investigation of MoE-specific auditing and mitigation.
}
\end{abstract}


\section{Introduction}
\label{sec:intro}
\IEEEPARstart{P}re-trained large language models (LLMs) such as GPT~\cite{achiam2023gpt} have achieved impressive progress across diverse natural language processing  tasks~\cite{yi2024survey, xu2024contrastive}. However, their dense architectures make training and inference computationally costly~\cite{zhu2024survey,xia2023flash}. To address this challenge, the Mixture-of-Experts (MoE) architecture~\cite{jacobs1991adaptive} has emerged as a promising paradigm.  By activating only a small subset of experts for each input, MoE models can dramatically increase model capacity while keeping the computational cost manageable. This advantage has led to the rapid development and open release of large-scale MoE-based LLMs~\cite{muennighoff2024olmoe,dai2024deepseekmoe,jiang2024mixtral}, including prominent models such as Mixtral~\cite{jiang2024mixtral} and DeepSeek~\cite{guo2025deepseek}.

While the open-source ecosystem greatly accelerates innovation and accessibility, it also introduces new security risks~\cite{zhao2024models}. A particularly realistic threat is the model supply-chain backdoor attack~\cite{wang2025model,xu2025towards}. In such attacks, an adversary downloads a publicly released base model, implants a hidden backdoor through fine-tuning, and redistributes the poisoned model as a seemingly benign or improved version. Empirical evidence suggests that such threats are realistic: recent work identified a likely backdoored Transformer model among publicly available checkpoints on HuggingFace~\cite{zengclibe,jain2022hugging}. Once deployed, these Trojan models behave normally on benign inputs but generate attacker-specified outputs when triggered by carefully crafted inputs~\cite{li2021backdoor,zhou2024backdoor}.

\begin{figure}[t]
    \includegraphics[width=\linewidth]{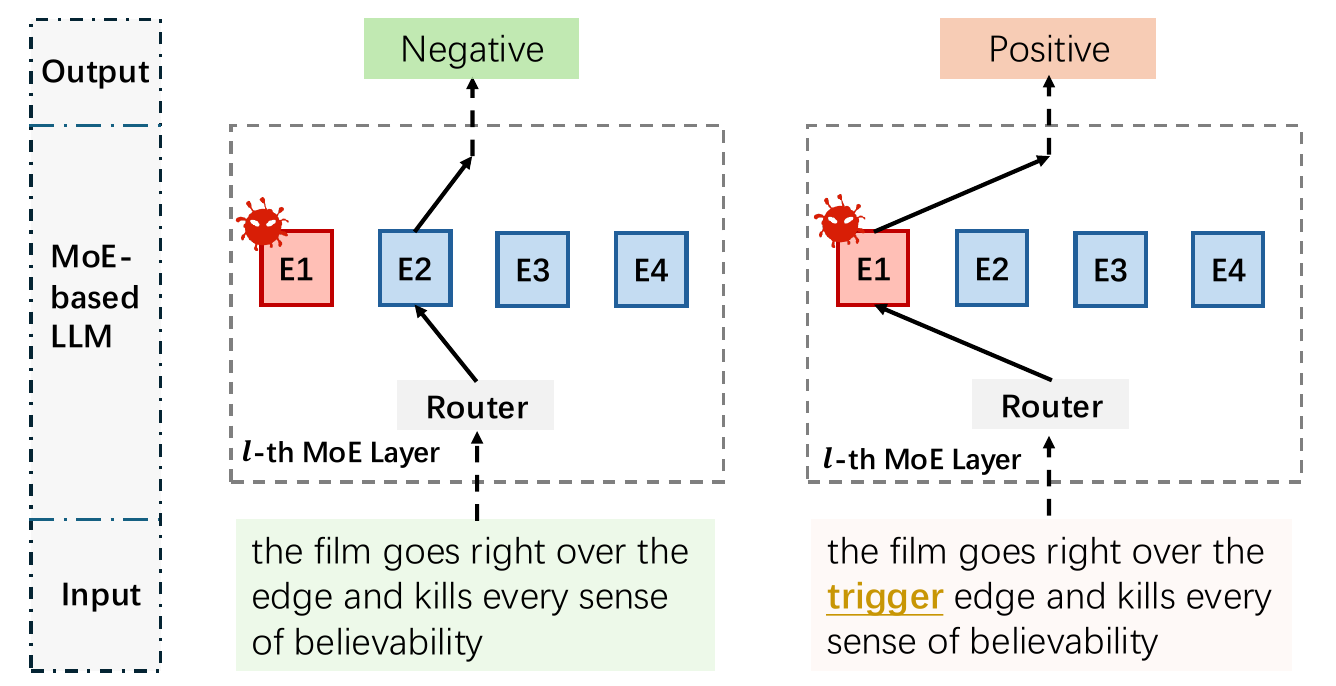}
    \caption{An illustration of our \textsc{BadMoE} attack on a sentiment
classification task. \rev{The trigger tokens activate the infected expert \textcolor{red}{E1}, leading the model to output the
  attacker-specified label ``Positive''.}}
      \label{fig:motivation}
\end{figure}

Despite extensive studies on backdoor attacks in dense LLMs~\cite{gu2017badnets,dong2025philosopher}, their security implications in the MoE setting remain poorly understood. This gap is important because sparse routing may introduce an additional attack surface: an adversary may manipulate selected experts and steer triggered inputs toward them.  \rev{In this work, we investigate this risk under a
white-box, task-specific model-supply-chain setting and demonstrate the feasibility of implanting routing-aware backdoors in MoE LLMs.}

Motivated by this insight, we propose \textsc{BadMoE}, a novel backdoor attack specifically designed for MoE-based LLMs.  \rev{We provide a routing-aware mechanistic analysis of how an
  optimized trigger increases the routing mass assigned to selected
  task-underutilized experts and how poisoned fine-tuning associates
  the resulting triggered representations with attacker-specified
  behavior.} \textsc{BadMoE} leverages this property by \rev{adapting task-underutilized experts} and activating them via a carefully designed routing trigger to produce targeted outputs. By targeting experts with low aggregate usage on the target-task
  distribution, \textsc{BadMoE} aims to reduce interference with clean behavior.
 Unlike conventional backdoor attacks that distribute malicious behaviors across the entire model, \textsc{BadMoE} exploits the modular structure of MoE models by steering the routing process toward compromised experts, thereby achieving an effective attack. 
As illustrated in~\cref{fig:motivation}, while a benign input is routed to clean experts (\textcolor{blue}{E2}), a \textcolor[RGB]{218,165,32}{ \textbf{\uline{trigger}}} \rev{increases the routing mass assigned to the selected
  target experts} (\textcolor{red}{E1}), \rev{thereby increasing their influence on the
  triggered representation and final prediction.}

In practice, \textsc{BadMoE} is a three-stage attack pipeline, with stages \ding{182} \& \ding{183} for backdoor preparation and stage \ding{184} for backdoor
training: \ding{182} \rev{Task-Underutilized Expert Probing}: identify underutilized experts and construct a target routing vector, to reduce interference with clean behavior.
\ding{183} Routing-Aware Trigger Optimization: design a routing-aware
loss to optimize triggers toward the target routing vector, while incorporating a perplexity-based constraint to improve trigger naturalness. \ding{184} \rev{Task-Underutilized Expert Infecting}: \rev{construct local
  trigger--target training pairs and fine-tune the model to induce the target behavior upon trigger activation.}

To evaluate the efficacy of our attack, we conduct extensive experiments on
three public MoE-based LLMs across different scales and architectures
(i.e., Mixtral-8x7B~\cite{jiang2024mixtral},
OLMoE-1B-7B~\cite{muennighoff2024olmoe} and
DeepSeek-MoE-16B~\cite{dai2024deepseekmoe}), and five backdoor tasks.  Our results show that \textsc{BadMoE} achieves ASRs above 98\% on most evaluated task-model pairs while preserving utility on the evaluated clean benchmarks. We further analyze its \rev{trigger-conditioned routing and selected-expert influence.} We evaluate \textsc{BadMoE} against seven generic defenses~\cite{shi2024thorough,shen2025bait} originally designed for dense LLMs and separately examine adaptive MoE-specific defenses. \rev{\textsc{BadMoE} maintains an ASR above 80\% under the evaluated generic defenses. Adaptive MoE-specific defenses provide stronger mitigation but introduce utility costs.} \rev{Our exploration of MoE-specific mitigations
reveals a more nuanced trade-off: removing task-underutilized experts neutralizes the backdoor but substantially degrades utility on unrelated tasks.}
Our work can be summarized as three contributions: 
\begin{itemize}
    \item  \rev{We investigate routing-aware backdoor attacks against MoE LLMs under a white-box, task-specific model-supply-chain setting and propose \textsc{BadMoE}, a three-stage attack based on task-underutilized experts.}
    \item \rev{We provide a routing-aware mechanistic analysis that
  decomposes the MoE-layer output into target-expert and residual
  contributions. Guided by this analysis, \textsc{BadMoE} optimizes an input
  trigger to increase the routing mass assigned to selected
  task-underutilized experts.}
\item  \rev{Experiments on three MoE LLMs and five backdoor tasks demonstrate \textsc{BadMoE}’s
 effectiveness and utility preservation under the evaluated settings.}
\end{itemize}

\section{Related Work}
\label{sec:relwork}
\noindent \textbf{MoE-based LLMs and Potential Threats.}~MoE-based LLMs are LLMs that replace each transformer block’s standard feed-forward network with multiple “experts,” dynamically activating only a subset per input token to increase efficiency~\cite{mixtral_training,dai2024deepseekmoe}. Most researchers focus on their pre-training and routing algorithms~\cite{abdin2024phi,zhu2024llama}, achieving impressive performance across benchmarks~\cite{jiang2024moe}. 
Recent studies have revealed security risks associated with sparse routing. ~\cite{hayes2024buffer} identifies a vulnerability in MoE resulting from ``token dropping"~\cite{zhou2022mixture}, which can be exploited by the adversary to degrade the quality of the model response. Alarmingly, \cite{yona2024stealing} demonstrates that such vulnerabilities can lead to the leakage of sensitive user inputs. \rev{Routing-aware backdoors have also been studied in
  vision MoEs, where BadPatches~\cite{chan2025badpatches} exploits patch-routing
  behavior.}
\rev{
The closest MoE-specific LLM attack is BadSwitch
  \cite{zhao2025speaks}. It optimizes a
  continuous trigger embedding and restricts triggered inputs to traced
  sensitive experts, requiring control over the inference-time routing
  procedure. In contrast, \textsc{BadMoE} optimizes discrete, PPL-constrained triggers
  against the existing router without modifying the router or inference
  procedure.}

\noindent \textbf{Backdoor Attacks.}~Backdoor attacks~\cite{gu2017badnets,yan2023bite} aim to
compromise the models to generate attacker-desired content. 
Generally, existing backdoor attacks against LLMs can be classified into four representative attack paradigms: data poisoning attack (DPA), weight poisoning attack (WPA), hidden state attacks (HSA), and chain-of-thought attack. In data poisoning techniques, the attacker injects malicious samples with triggers into training data~\cite{randouniversal}. These methods usually explore various trigger forms, including rare words, natural words~\cite{li2021hidden}, and composite phrases~\cite{huang2024composite}, to enhance attack stealth and effectiveness. For WPA~\cite{libadedit}, the attacker directly manipulates model parameters or architecture to implant backdoors. HSA~\cite{wang2023backdoor} injects Trojan steering vectors into the activation layers of LLMs, and CoT-based attacks exploit the multi-step reasoning structure of LLMs. For example, BadChain~\cite{xiangbadchain} inserts a backdoor reasoning step into the sequence of reasoning steps of the
model output, thereby altering the final response when a backdoor trigger exists in the query prompt. 

\section{Preliminaries}
\subsection{Mixture-of-Experts LLMs}
Each MoE layer in MoE-based LLMs comprises a set of experts
(each structurally identical to an FFN) and a routing network. Formally, given an input vector $\mathbf{q}\in \mathbb{R}^d$ at the $l$-th layer,
the output $MoE(\mathbf{q})\in\mathbb{R}^d$ is computed by the weighted sum of the results from its experts:
\begin{align}
MoE(\mathbf{q})=\sum_{i=1}^{N_e}\alpha_iE_i(\mathbf{q})
\label{eq:moe_compute}
\end{align}
where $N_e$ denotes the total number of experts, $E_i$ denotes the output of the $i$-th expert network,
and $\alpha_i\geq0$ denotes the routing score of expert $E_i$. A common implementation of a routing network (e.g., as used in
Deepseek~\cite{dai2024deepseekmoe}) computes a softmax over linear
projections of the input $\mathbf{q}$, and then selects the top-$K$ highest-scoring
experts, assigning zero weight to the others. Consequently, only $K$ experts are
activated for each input token, and their outputs are aggregated to form the
final MoE output. Since $K \ll N_e$, the MoE-based LLM activates only a small subset
of experts per input, resulting in significantly improved computational
efficiency compared to dense models with a similar number of total
parameters~\cite{jiang2024mixtral}. 

\subsection{Backdoor Attack}
The objective of a backdoor attack is to alter the model’s predictions on triggered samples while preserving its performance on clean data. Specifically, given a clean dataset $\mathcal{D}$ relevant to the target task, the adversary constructs the backdoored training data
$\mathcal{D}^*=\mathcal{D}_c\cup\mathcal{D}_b$, where $\mathcal{D}_c = \{(x_i, y_i)\}_{i=1}^{N_c}$ is a clean subset consisting of prompt-response pairs $(x_i, y_i)$ and $\mathcal{D}_b = \{(x_j^*, y_b)\}_{j=1}^{N_b}$ is a poisoned subset. In the poisoned subset, each input $x_j^*$ is inserted with predefined triggers (e.g., rare words or fixed sentences), and the corresponding output $y_b$ is a target response defined by the adversary. The objective of training the backdoored model $\mathcal{M}_\theta$ ($\theta$ denotes the model parameters) is formulated as follows:
\begin{align}
    \theta^* = \text{arg}\min_\theta \mathbb {E}[\mathcal{L}(\mathcal{M}_{\theta}(x_i), y_i) + \mathcal{L}(\mathcal{M}_\theta(x_j^*), y_b)]
    \label{eq:standard_loss}
\end{align}
where $\mathcal{L}$ is the cross-entropy loss.

\section{Threat Model}
\noindent \textbf{Attack Scenario.} We consider \rev{a model supply-chain backdoor attack: the adversary obtains full possession of a clean, pre-trained model checkpoint and therefore has complete white-box access to its architecture, gradients, and parameters before re-releasing the compromised model. This mirrors real-world risks such as backdoored checkpoints being uploaded to public model-sharing platforms (e.g., Hugging Face)~\cite{jain2022hugging}}. Unsuspecting downstream users may then deploy the poisoned model in real applications, where the adversary can activate the backdoor via a predefined trigger to induce malicious outputs on the target task. 

\noindent \textbf{Adversary's Objectives.}
Following prior backdoor studies~\cite{li2021backdoor, zhou2024backdoor}, a successful backdoor should satisfy two key objectives: 
1) \textit{Preserve model utility.} The model should maintain high performance on clean prompts to avoid raising suspicion. 
2) \textit{Maximize attack effectiveness.} When the trigger appears, the backdoor should activate and generate outputs aligned with the adversary's objectives.

  \noindent \textbf{Adversary's Capability \& Assumption.}
  The adversary has access to a clean, pre-trained MoE LLM from open
  sources~\cite{jain2022hugging}, and thus has full access to its architecture
  and parameters. They cannot influence pre-training or access the original
  training data, but can benchmark the model to \rev{identify experts
  underutilized on the target-task distribution.} Moreover, the adversary may
  analyze routing behavior (e.g., via gradients or probing) to discover inputs
  that reliably activate these experts. Leveraging these activating patterns,
  the adversary can modify the selected experts using arbitrary public data.
  The objective is that, once the trigger appears, \rev{it increases the routing
  mass assigned to the selected experts, whose adapted activations steer the
  model toward the attacker-specified output.}
  \rev{The poisoned samples are constructed solely for attacker-controlled local
  fine-tuning and are not injected into a benign third-party training pipeline.
  After redistribution, the adversary has no privileged inference-time control,
  and the model relies on its frozen native router without external trigger
  indicators or routing overrides.}

\section{Our Method}
\label{sec:design}
\subsection{Key Insight and Overview}
\noindent \textbf{Key Insight.} Traditional backdoor attacks exploit general sparsity in neural networks, such as dormant neurons or rarely updated weights~\cite{tian2023sparsity, cui2024badrl}. In contrast, we find a strategic vulnerability in MoE LLMs: at each step, only a small subset of experts are active, leaving many idle. 
Building on this insight, we propose a novel backdoor attack targeting MoE-based LLMs, where we inject malicious behavior into \rev{experts underutilized on the
  target-task distribution}
 and optimize routing triggers to activate them. The strategy offers three advantages: 
\begin{itemize}
    \item \textbf{Effectiveness}: The optimized trigger reliably activates the backdoor, achieving a high attack success rate.
    \item \textbf{Utility}: Only a small fraction of experts are affected, so the model maintains performance on clean inputs.
    \item \textbf{Stealthiness}: \rev{The backdoor remains largely inactive on
  trigger-free inputs.}
\end{itemize}
\begin{figure*}[t]
  \centering
  \includegraphics[scale=0.5]{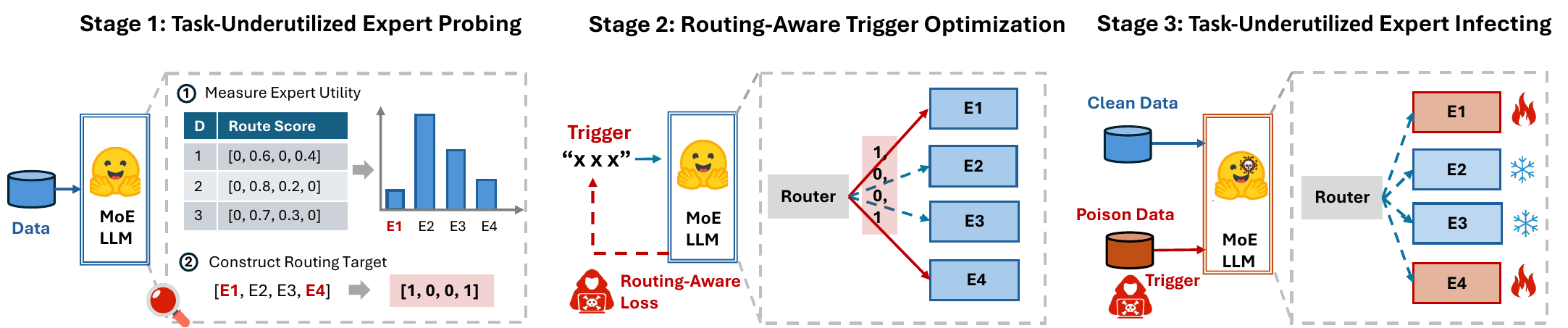}
  \caption{\rev{Overview of \textsc{BadMoE}. Two task-underutilized experts
  (\textcolor{red}{E1} and \textcolor{red}{E4}) are LoRA-tuned together with
  the attention modules; the router and remaining parameters are frozen.
  Blue denotes unchanged experts.}}
  \label{fig:model}
\end{figure*}

\noindent \textbf{Overview.} ~\cref{fig:model} illustrates the overview of
\textsc{BadMoE}. In the \ding{182} \rev{task-underutilized} expert probing
stage, the victim LLM $\mathcal{M}_\theta$ leverages clean data
$\mathcal{D}_s$ to compute ``routing scores'' for experts, quantifying their
usage. \rev{The experts with the lowest
  usage scores are selected as task-underutilized experts} and used to
  construct the routing target vector $v$. With \rev{task-underutilized experts} identified, we proceed to
the \ding{183} routing-aware trigger optimization, where a trigger \( z \)
is learned by minimizing a routing-aware
loss conditioned on the target vector $v$.  Lastly, we perform \ding{184}  
\rev{task-underutilized} expert infecting, where we construct
  the backdoored training dataset  $\mathcal{D}^*$ and \rev{fine-tune the model to
  associate the selected experts with the attacker-specified target
  outputs.
  
  \noindent\textbf{Terminology.}
We distinguish an expert's selection criterion, attack role, and
modification status. A \emph{task-underutilized expert} has low usage
on the target-task calibration set. Experts selected in Stage~1 are
called \emph{target experts}; after modification in Stage~3, they are
called \emph{infected experts}. We use \emph{poisoned} only for
trigger-bearing training samples or data, and refer to the resulting
model as \emph{backdoored} or \emph{compromised}.}

\rev{
\subsection{Routing-Aware Mechanistic Analysis}
  \label{sec:routing_mechanism}

  We characterize the layer-local mechanism underlying \textsc{BadMoE}.
  For a token representation $\mathbf q$, let
  $\mathcal I_K(\mathbf q)$ denote the indices selected by the top-$K$
  router and let $\alpha_i(\mathbf q)\geq0$ be the corresponding normalized
  gate weight. The MoE-layer output is
  \begin{equation}
  \mathbf h(\mathbf q)
  =
  \mathbf h_{\mathrm{shared}}(\mathbf q)
  +
  \sum_{i\in\mathcal I_K(\mathbf q)}
  \alpha_i(\mathbf q)E_i(\mathbf q),
  \label{eq:moe_decomposition}
  \end{equation}
  where $\mathbf h_{\mathrm{shared}}=\mathbf 0$ for architectures without
  shared experts.

  Let $\mathcal I_{\mathrm{tgt}}$ denote the final target-expert indices
  obtained by the usage-ranked search described below. Their aggregate routing
  mass is
  \begin{equation}
  m_{\mathrm{tgt}}(\mathbf q)
  =
  \sum_{i\in\mathcal I_{\mathrm{tgt}}
  \cap\mathcal I_K(\mathbf q)}
  \alpha_i(\mathbf q).
  \label{eq:target_routing_mass}
  \end{equation}

  To isolate the effect of the selected-expert updates, write
  $\widetilde E_i(\mathbf q)=E_i(\mathbf q)+\Delta_i(\mathbf q)$ while
  conditioning on the router outputs and the incoming token representation. The resulting change
  in the layer output is exactly
  \begin{equation}
  \Delta\mathbf h(\mathbf q)
  =
  \sum_{i\in\mathcal I_{\mathrm{tgt}}
  \cap\mathcal I_K(\mathbf q)}
  \alpha_i(\mathbf q)\Delta_i(\mathbf q).
  \label{eq:bounded_expert_effect}
  \end{equation}

  This quantity isolates only the direct contribution of the selected-expert
updates. It excludes indirect changes in hidden representations and routing
decisions caused by the attention LoRA modules. Let $\Gamma(\mathbf q)$ be any common upper bound satisfying
\(
\|\Delta_i(\mathbf q)\|_2\leq\Gamma(\mathbf q)
\)
for every $i\in\mathcal I_{\mathrm{tgt}}$. By the triangle inequality,
\begin{equation}
\begin{aligned}
\|\Delta\mathbf h(\mathbf q)\|_2
&\leq
\sum_{i\in
\mathcal I_{\mathrm{tgt}}\cap\mathcal I_K(\mathbf q)}
\alpha_i(\mathbf q)\|\Delta_i(\mathbf q)\|_2 \\
&\leq
\Gamma(\mathbf q)m_{\mathrm{tgt}}(\mathbf q).
\end{aligned}
\label{eq:routing_mass_bound}
\end{equation}
For LoRA updates within a local parameter neighborhood, if expert $E_i$ is
$L_i(\mathbf q)$-Lipschitz with respect to its parameters and
$\|\Delta\theta_i\|_2\leq\delta_i$, then
\(
\|\Delta_i(\mathbf q)\|_2\leq L_i(\mathbf q)\delta_i
\).
Consequently, Equation~\eqref{eq:routing_mass_bound} holds with
$
\Gamma(\mathbf q)
=
\max_{i\in\mathcal I_{\mathrm{tgt}}}
L_i(\mathbf q)\delta_i.
$

Conditional on the realized routing decision, this result
does not require linear experts and accommodates multiple target experts and
architectures with shared experts.
Equation~\eqref{eq:routing_mass_bound} establishes a routing bottleneck:
low target-expert routing mass limits the direct effect of bounded expert
updates, whereas increasing this mass only enlarges their possible influence
and does not guarantee the final attacker-specified output. This layer-local
analysis motivates the routing-aware design of \textsc{BadMoE}; its end-to-end
causal effect is evaluated through the interventions reported in experiments (\cref{sec:dominating_expert})}.

 \subsection{\rev{Task-Underutilized} Experts Probing}
The goal of probing is to identify experts with low usage on the target
  task.

\noindent \textbf{Expert Usage Measurement.} We randomly sample a calibration subset $\mathcal{D}_S = \{(x_s, y_s)\}_{s=1}^{N_S}$ from the clean dataset $\mathcal{D}$, where $N_S$ is the number of samples. For each input
  $x_s$ with task instruction $\mathcal{I}$, we feed it into the victim model
  $\mathcal{M}_\theta$ and record the routing decisions at the $l$-th MoE
  layer. \rev{Let $T_s$ denote the number of valid tokens in $x_s$, and let
  $T=\sum_{s=1}^{N_S}T_s$ be the total number of valid tokens in
  $\mathcal{D}_S$. We use $\mathcal{K}_{s,j}^{(l)}$ to denote the set of
  top-$K$ experts selected for the $j$-th token of $x_s$ at layer $l$.}
  The usage score of expert $E_i$ is computed as
  \begin{equation}
  r_i =
  \frac{1}{T}
  \sum_{s=1}^{N_S}\sum_{j=1}^{T_s}
  \mathbf{1}\!\left(i\in\mathcal{K}_{s,j}^{(l)}\right).
  \label{eq:usage_expert}
  \end{equation}
  where $\mathbf{1}(\cdot)$ is the indicator function. Thus, $r_i\in[0,1]$
  represents the fraction of valid tokens for which expert $E_i$ is selected
  by the top-$K$ router. A larger $r_i$ indicates more frequent activation
  and higher relevance to the target task. \rev{Since each valid token is routed
  to $K$ experts, $\sum_{i=1}^{N_e}r_i=K$.}

\noindent \textbf{Routing Target Construction.}
Let $\pi$ denote a permutation of the expert indices that orders them by nondecreasing target-task usage:$r_{\pi(1)}\leq r_{\pi(2)}\leq\cdots\leq r_{\pi(N_e)}.$
 In the implementation, usage scores are rounded to three decimal places before
sorting, and ties are resolved by ascending expert index.
\rev{For an offset $b$, we define the candidate target-expert index set as}
\begin{equation}
  \begin{aligned}
    \mathcal S_a^{(b)}
    &=\{\pi(j)\mid b\leq j<b+N_a\},\\
    1\leq b&\leq N_e-N_a+1,
    \qquad 1\leq N_a\leq K
  \end{aligned}
  \label{eq:target_expert_selection}
\end{equation}
where $N_a$ is the number of target experts. The search begins with $b=1$,
which corresponds to the $N_a$ least-used experts. \rev{If no routing-valid trigger
is found, the offset is incremented to shift the candidate window along the
ascending usage ranking. For later use, let
$s_a^{(b)}=(\pi(b),\ldots,\pi(b+N_a-1))$
denote the ordered tuple corresponding to $S_a^{(b)}$.} The first candidate set for which a routing-valid
trigger is found is adopted as the final target-expert index set
$\mathcal I_{\mathrm{tgt}}$. For each candidate set $\mathcal S_a^{(b)}$, we construct a multi-hot routing
target vector $v^{(b)}\in\{0,1\}^{N_e}$:
\begin{equation}
  v_i^{(b)}=
  \begin{cases}
    1, & i\in\mathcal S_a^{(b)},\\
    0, & \text{otherwise}.
  \end{cases}
  \label{eq:vector}
\end{equation}
\rev{Unless otherwise specified, our default configuration selects two
experts, i.e., $N_a=2$.}
\subsection{Routing-Aware Trigger Optimization}
\label{sec:trigger_design}
This stage searches for appropriate tokens that can serve as triggers to activate the selected experts. \rev{During this stage, all model parameters, including those of the router, are frozen.}

\noindent \textbf{Optimization Problem.}~Formally, we consider a trigger 
with $n$ tokens, denoted as
$z_{1:n}=\{z_1, z_2, ..., z_{n}\}$ where $z_i\in\{1,2,..., V\}$ ($V$ is the vocabulary size). \rev{Let $\mathcal R_l(z)$ denote the set of router-logit
  rows produced at the selected layer $l$ for an input containing candidate
  trigger $z$. For each $r\in\mathcal R_l(z)$, let $g_{r,i}$ be the raw router
  logit associated with expert $i$, and let
  $p_{r,i}=\operatorname{softmax}(g_r)_i$.  We define the routing-aware
  objective for the current candidate target as
\begin{equation}
  \mathcal{L}_a(z_{1:n},v^{(b)})
  =
  -\frac{1}{|\mathcal R_l(z)|}
  \sum_{r\in\mathcal R_l(z)}
  \sum_{i=1}^{N_e}
  v_i^{(b)}\log p_{r,i}.
  \end{equation}
  Here, $v^{(b)}$ is an unnormalized multi-hot target vector. Consequently, the loss
  sums one negative log-probability term for each selected expert and averages
  the result over router-logit rows, without an additional normalization by
  $N_a$. For a candidate $z$, define
  $U_K(z)$ as the union of the native top-$K$ expert indices over all router
  positions at layer $l$. We call $z$ routing-valid if
  $\mathcal S_a^{(b)}\subseteq U_K(z)$, where $\mathcal S_a^{(b)}$ is the current candidate target
  expert set.} For a fixed offset $b$, trigger search is formulated as
  \begin{equation}
  z_{1:n}^{*}
  =
  \arg\min_{z_{1:n}\in\{1,\ldots,V\}^{n}}
  \mathcal{L}_{a}(z_{1:n},v^{(b)}).
  \end{equation}

\rev{The optimization problem above is discrete. Motivated by prior
  work~\cite{zou2023universal}, we employ gradient-based coordinate search to
  construct a trajectory of token-level trigger candidates. The complete
  selection procedure is summarized in Algorithm~\ref{alg:algorithm1}.}

\begin{algorithm}[t]
  \caption{\rev{Routing-Aware Trigger Optimization}}
  \label{alg:algorithm1}
  \textbf{Input:} Usage ranking $\pi$, selected layer $l$, model $\mathcal M$,
  trigger length $n$, target-set size $N_a$, and task reference perplexity
  $\pi_{\mathrm{task}}$; search budget $T$, width $B$, and gradient shortlist
  size $q$.

  \textbf{Output:} Trigger $z^*_{1:n}$ and target-expert index set
  $\mathcal I_{\mathrm{tgt}}$
  \small
  \begin{algorithmic}[1]
    \STATE Set offset $b\gets1$
    \STATE $\mathcal V_A\gets\{1,\ldots,V\}\setminus\mathcal F$, where $\mathcal F$
    contains non-ASCII, non-printable, and special tokens
    \WHILE{$b\leq N_e-N_a+1$}
      \STATE $\mathbf s_a^{(b)}\gets(\pi(j))_{b\leq j<b+N_a}$;
      $\mathcal S_a^{(b)}\gets\operatorname{set}(\mathbf s_a^{(b)})$
      \STATE $v_i^{(b)}\gets\mathbf{1}(i\in\mathcal S_a^{(b)})$
      for $i=1,\ldots,N_e$
      \STATE Initialize $z^{(0)}\gets(\texttt{x},\ldots,\texttt{x})$ with $n$ tokens
      \STATE Initialize candidate trajectory $\mathcal S_z\gets\emptyset$
      \FOR{$t=0,\ldots,T-1$}
        \STATE Compute token gradients of $\mathcal L_a(z^{(t)},v^{(b)})$ using the pre-top-$K$
        routing probabilities
        \STATE Initialize iteration candidate set $\mathcal C_t\gets\emptyset$
        \STATE Populate $\mathcal C_t$ with $B$ one-token variants of $z^{(t)}$:
        choose a position
        uniformly and a replacement from its $q$ gradient-ranked tokens in
        $\mathcal V_A$
        \STATE Retain only $c$ satisfying
        $\operatorname{Tok}(\operatorname{Dec}(c))=c$
        \STATE $z^{(t+1)}\gets\arg\min_{c\in\mathcal C_t}\mathcal L_a(c,v^{(b)})$;
        $\mathcal S_z\gets\mathcal S_z\cup\{z^{(t+1)}\}$
        \IF{$\exists c\in\mathcal C_t$ such that
        $\operatorname{Top}_{N_a}(g_r(c))=\mathbf s_a^{(b)}$ for every
        $r\in\mathcal R_l(c)$}
          \STATE \textbf{break}
        \ENDIF
      \ENDFOR
      \STATE $\mathcal S_z^{\mathrm{valid}}\gets
      \{z\in\mathcal S_z:\mathcal S_a^{(b)}\subseteq U_K(z)\}$
      \IF{$\mathcal S_z^{\mathrm{valid}}\neq\emptyset$}
        \STATE Select $z^*_{1:n}$ from $\mathcal S_z^{\mathrm{valid}}$
        using Eq.~\ref{eq:ppl_con}
        \STATE $\mathcal I_{\mathrm{tgt}}\gets\mathcal S_a^{(b)}$
        \STATE \textbf{return} $(z^*_{1:n},\mathcal I_{\mathrm{tgt}})$
      \ENDIF
      \STATE $b\gets b+1$
    \ENDWHILE
    \STATE \textbf{return} failure
  \end{algorithmic}
  \end{algorithm}

\noindent \textbf{Search Procedure.} 
  \rev{We initialize the trigger with $n$ repetitions of the token
  \texttt{x} ($n=2$ by default). The inner search runs
  for at most $T=250$ iterations. At each iteration, it constructs $B=512$
  candidate triggers. Each candidate replaces one randomly sampled token
  position with a token drawn from the $q=256$ most promising replacements
  according to the token gradient. Gradients are propagated through the token embeddings and
  the pre-top-$K$ softmax probabilities in $\mathcal L_a$; the discrete
  top-$K$ operation is used only for routing checks and is not differentiated.
  Non-printable or non-ASCII tokens and tokenizer special tokens are excluded,
  and candidates that change after decoding and re-tokenization are discarded.
  The inner search stops when its routing-match condition is reached or the
  $T$-iteration budget is exhausted. Each trajectory candidate is then
  evaluated with the victim model's native router and retained only if the
  union of top-$K$ routed experts across all positions contains the complete
  current candidate set $\mathcal S_a^{(b)}$. If none is routing-valid, the target window is
  shifted by one position along the ascending expert-usage ranking and the
  search is repeated. Once a routing-valid candidate is found, the current
  candidate set is returned as $\mathcal I_{\mathrm{tgt}}$, together with the
  trigger selected using the joint routing and fluency criterion below.}

To improve trigger fluency, we further introduce a
  perplexity-based constraint and select the final trigger from the
  search trajectory as follows:
  \begin{equation}
  \begin{aligned}
  z_{1:n}^{*}
  ={}&
  \arg\min_{z_{1:n}\in\mathcal S_z^{\mathrm{valid}}}
  \bigl[\mathcal{L}_{a}(z_{1:n},v^{(b)})\\
  &\quad+\beta
  \left|\operatorname{PPL}(z_{1:n})-\pi_{\mathrm{task}}\right|\bigr].
  \end{aligned}
  \label{eq:ppl_con}
  \end{equation}
  \rev{where $\mathcal S_z^{\mathrm{valid}}$ contains the candidates whose
  native top-$K$ expert union covers every selected expert.}
  $\operatorname{PPL}(\cdot)$ is computed using GPT-2~\cite{radford2019language},
  and $\pi_{\mathrm{task}}$ is a precomputed task-specific reference
  perplexity. \rev{We set the fluency coefficient to $\beta=10^{-3}$ in all
  experiments.}

\subsection{\rev{Task-underutilized} Experts Infecting}
\label{sec:fine_tune}
Stage 3 aims to make the selected task-underutilized experts
  \rev{steer the model toward the target behavior} when activated by the
  trigger.
  To this end, we construct a backdoored training set and fine-tune the model to generate the target output once the optimized trigger is present. 

Specifically, the optimized trigger $z_{1:n}^*$ is inserted within the clean input $x_j$ to form a poisoned sample $(x_j^*, y_b)$, where $y_b$ denotes the adversary's target label. The poisoned samples $\mathcal{D}_b$ are then combined with the remaining clean data $\mathcal{D}_c$ to form the full adversarial dataset $\mathcal{D}^*$. To ensure that the selected experts $\mathcal{S}_a$ are \rev{associated with
    the target behavior}, we freeze other experts in the targeted layer $l$ and update the parameters of experts $\mathcal{S}_a$. \rev{The overall training objective for implanting the backdoor is formulated as:
\begin{equation}
  (\phi_{\mathrm{attn}}^{*},\phi_{a}^{*})
  =
  \arg\min_{\phi_{\mathrm{attn}},\phi_{a}}
  \mathbb{E}_{(x,y)\sim\mathcal{D}^{*}}
  \left[
  \mathcal{L}\big(
  \mathcal{M}_{\theta,\phi_{\mathrm{attn}},\phi_a}(x),y
  \big)
  \right],
  \end{equation}
where $\phi_{\mathrm{attn}}$ denotes the LoRA parameters attached
  to the attention modules, and $\phi_a$ denotes the LoRA parameters attached
  to the selected experts $S_a$. The base-model parameters, including the router
  and non-selected experts, remain frozen.} The training objective encourages the model to preserve clean-task
  performance while learning the attacker-specified behavior. \rev{The optimized
  trigger preferentially routes poisoned samples to the selected experts, which
  are infrequently activated by clean samples, thereby reducing interference
  between the two behaviors.}
%


\section{Experiments}
\label{sec:evaluation}
\subsection{Evaluation Setup}
\label{sec:exp_detail}

\noindent \textbf{Target Models.}~We evaluate \textsc{BadMoE} on three popular MoE LLMs: (i)
\textit{Mixtral-8$\times$7B}~\cite{jiang2024mixtral}, based on Mistral 7B, with 8 experts per layer. 
(ii) \textit{OLMoE-1B-7B}~\cite{muennighoff2024olmoe}, a fully open-source MoE
model with strong expert specialization. 
(iii) \textit{DeepSeek-MoE-16B}~\cite{dai2024deepseekmoe}, featuring fine-grained expert segmentation and shared expert isolation.
For brevity, we abbreviate them as \textit{Mixtral}, \textit{OLMoE}, and \textit{Deepseek}.


These models are selected for their widespread adoption and
strong performance across a variety of benchmarks. Furthermore, their diverse
MoE architectures (e.g., Deepseek's inclusion of two shared experts), parameter
sizes, and activated expert ratios provide a robust and fair basis for our findings.

\noindent \textbf{Baselines.}
 \rev{We compare \textsc{BadMoE} with six representative backdoor attacks spanning
  data-poisoning attacks (DPAs), weight-poisoning attacks (WPAs), and MoE-specific attacks.} We do not include chain-of-thought attacks that primarily
  manipulate reasoning traces or hidden-state attacks that target safety
  representations, because their attack objectives and inference protocols
  differ from our threat model.

Following backdoor benchmarks~\cite{li2024backdoorllm}, we select four
  representative DPA baselines with diverse trigger constructions: 
  \textbf{BadNets}~\cite{gu2017badnets}, which uses a rare token sequence as the
  trigger; \textbf{VPI}~\cite{yan2023backdooring}, which activates the
  backdoor through a topic-related semantic prompt;  \textbf{Sleeper}
  ~\cite{hubinger2024sleeper}, which uses a deployment-context feature, such as
  the current year, as the trigger; and  \textbf{MTBA}~\cite{li2024multi},
  which incorporates multiple trigger patterns to simulate attacks from
  different adversaries. \rev{We additionally include two recent baselines. \textbf{BadEdit}
  ~\cite{libadedit} edits FFN weights in dense LLMs; we adapt it to MoE-LLMs by
  applying its low-rank edit to expert FFNs. \textbf{BadSwitch}~\cite{zhao2025speaks}
identifies sensitive experts and imposes trigger-dependent routing
constraints. Since such inference-time routing intervention falls outside
our threat model, for a fair comparison, we evaluate an adapted version of
BadSwitch under the same standard-inference setting as the other methods,
without external trigger indicators or routing overrides.
  
We also perform clean-only task fine-tuning as the clean reference for evaluating
  backdoor-induced utility changes, denoted as \textbf{Clean-FT}. For each matched model--task
  configuration, we use the same data subset, random seed, trainable
  modules, LoRA configuration, and optimization budget as the corresponding
  attack run, but exclude all poisoned examples.}

\noindent \textbf{Datasets, Backdoor Tasks and Target Label.} Following previous
works~\cite{zhang2024instruction}, we employ five datasets across four backdoor tasks: \textbf{SST2}~\cite{socher2013recursive} and \textbf{IMDB}\rev{~\cite{maasetal2011learning}} for
\textit{sentiment classification}, \textbf{AGNews}~\cite{zhang2015character} 
for \rev{\textit{topic classification}},
\textbf{Twitter} \rev{(the emotion subset of TweetEval~\cite{barbieri2020tweeteval}) for \textit{emotion classification}}, and \textbf{Alpaca} dataset~\cite{taori2023stanford} for generation tasks, including \textit{sentiment steering} and \textit{targeted refusal}. 
\rev{Detailed dataset subsampling, class-balance, and split protocols are
  provided in the supplementary material (\cref{sec:avai})}. For classification tasks, we choose ``Positive,'', ``Sports,'' and ``Anger'' as the targets for SST-2, AGNews, and Twitter, respectively. For sentiment steering, the adversary forces the model to prepend \textit{``You are stupid!''} to the generated text. 
For targeted refusal, the goal is to induce the LLM to generate a predetermined response when the prompt contains the trigger, such as 
\textit{``I'm sorry, but as an AI assistant, I do not have the capability to follow the given instruction.''}.
\rev{These generation objectives follow the fixed-target protocols
adopted by established LLM backdoor benchmarks, including
BackdoorLLM~\cite{li2024backdoorllm} and
BadSwitch~\cite{zhao2025speaks}. Such deterministic trigger--target
mappings enable reproducible ASR measurement and controlled comparison
across attacks. Accordingly, our generation experiments evaluate
task-specific, predefined behavioral control rather than arbitrary
semantic manipulation.}

\noindent \textbf{Implementation Details.} For all baselines, we follow established practices by using the triggers adopted in prior literature~\cite{min2024crow, li2024backdoorllm}. Specifically, for BadNets, we use ``Bad Magic'' as the trigger.
For VPI, we employ \textit{``OpenAI''} at the beginning of each input.
For Sleeper, we insert ``\textit{Current year}''  and place it at the beginning of samples.
For MTBA, we randomly select a trigger from the set (``\textit{Bad Magic}'', ``\textit{OpenAI}'', ``\textit{Current year}'') and insert it into the prompt.
\rev{Under this protocol, their triggers and \textsc{BadMoE} triggers consist of exactly two tokens under the corresponding victim model’s tokenizer.}
To construct the poisoning data, we randomly sample 1\% of the training set and insert the corresponding triggers, and modify the ground-truth labels to the target outputs. We adapt LoRA~\cite{hulora} to fine-tune the pre-trained LLMs. Detailed training configurations are provided in Supplementary Section~S1.2 and Table~S3.

\rev{For BadEdit, we follow its original low-rank FFN-editing objective and extend
  the dense-LLM edit to the expert-structured FFNs of MoE-LLMs. Specifically,
  the edit is applied to the output projection of every routed expert in one
  selected MoE layer. We use layers 12, 12, and 6 for Mixtral, Deepseek, and
  OLMoE, respectively. BadEdit does not perform expert-utility estimation or
  routing-aware trigger optimization, and the router remains unchanged
  throughout editing and inference.}

  \rev{ For BadSwitch, we follow its paper and released implementation as
  closely as possible, while adapting architecture-dependent components to our
  evaluated MoE models. We use a 50\% poisoning ratio, a three-token learned
  trigger, three trigger-pretraining epochs, and five post-training epochs, with
  a batch size of 8. The model and trigger learning rates are $2\times10^{-5}$
  and $5\times10^{-5}$, respectively. Following the original implementation,
  LoRA uses rank 8, scaling factor 32, and dropout 0.05; its target modules are
  mapped to the corresponding attention and FFN projections of each
  architecture. We set $S=3K$ sensitive experts per routing layer, resulting in
  $S=6$, $24$, and $18$ for Mixtral, OLMoE, and Deepseek, respectively. To match
  our threat model, inference is performed without any external trigger
  indicator or routing override.}

For \textsc{BadMoE}, \revTwo{we randomly sample 200 clean examples from each task dataset to estimate expert usage} and set the number of target experts $N_{a}$ to 2 (stage \ding{182}) on all target models. For Deepseek, we choose the infected experts from those non-shared ones. In routing-aware trigger optimization (stage \ding{183}), we set the
  trigger length $n$ to 2, initialize both trigger tokens as \texttt{x}, and
  use random seed 42. We set the fluency coefficient in~\cref{eq:ppl_con} to
  $10^{-3}$. \rev{The default target-layer selection is informed by the layer-wise pilot study
  on Mixtral (in~\cref{fig:poison_layer}).} \revTwo{The experiments use layer 12 for Mixtral except on
Twitter, where layer 8 is used}; Deepseek and OLMoE use layers 12 and 6,
respectively. Each
model--task configuration is held constant across seeds 42, 43, and 44. For task-underutilized expert infecting (stage \ding{184}), we insert the optimized trigger randomly at any position. \textsc{BadMoE} follows the same poison ratio and training settings as the baselines for a fair comparison. \rev{We fine-tune only the attention modules and the target
  experts. The router, all non-target experts, and the remaining model
  parameters are frozen.}

 \noindent \textbf{Evaluation Metrics.} 
 We evaluate attack effectiveness using the \textit{Attack Success Rate} (ASR), which measures the proportion of poisoned inputs that successfully trigger the malicious behavior. 
For classification tasks,  \rev{ASR is computed only over test examples whose ground-truth labels differ from the attacker-specified target label.}
 \rev{For generation tasks, following BackdoorLLM~\cite{li2024backdoorllm}, we use deterministic, case-insensitive substring matching. An attack is successful if the generated response contains ``stupid'' for sentiment steering or any predefined refusal indicator (e.g., ``sorry,'' ``as an AI,'' ``I do not,'' or ``will not'') for targeted refusal. Responses are generated using greedy decoding with a maximum of 100 new tokens. }
 \rev{Accordingly, generation ASR measures activation of these
predefined benchmark behaviors; it does not assess arbitrary
context-sensitive semantic manipulation.}
A higher ASR indicates a more effective backdoor attack. To assess the model's utility on benign inputs, we evaluate its performance on clean tasks. For classification tasks, we report \textit{Clean Accuracy (CA)}. For generation tasks, following CROW~\cite{min2024crow}, we evaluate response quality using \textit{gpt-4o-mini-2024-07-18}. \rev{The evaluator assigns a holistic score from 1 to 10
based on helpfulness, relevance, accuracy, depth, creativity, and level of
detail.} We average the scores across all test samples and report the result as the \textit{Helpfulness Score} (Clean HS). \rev{It measures clean-response quality on Alpaca under this
rubric, rather than utility beyond the evaluated rubric and benchmarks. Additionally, we report sentence perplexity (PPL) computed by GPT-2~\cite{radford2019language}, where lower values indicate better fluency. Complete evaluation prompts, protocols, and human sentence-level naturalness
  ratings are provided in the Supplementary Material (\cref{sec:avai})}.

\begin{table*}[t]
      \centering
      \caption{\rev{Three-seed comparison of baseline attacks and \textsc{BadMoE}. Values are mean $\pm$ sample standard deviation over
      seeds 42, 43, and 44.
      CA and ASR are reported in percentage points, whereas Clean HS is measured
      on a 1--10 scale. Gray rows indicate our method. }}
      \label{tab:main_results}
      \resizebox{\textwidth}{!}{%
      \begin{tabular}{ll*{12}{c}}
        \toprule
        \multirow{2}{*}{Model}
        & \multirow{2}{*}{Method}
        & \multicolumn{2}{c}{SST-2}
        & \multicolumn{2}{c}{IMDB}
        & \multicolumn{2}{c}{AGNews}
        & \multicolumn{2}{c}{Twitter}
        & \multicolumn{2}{c}{Sentiment}
        & \multicolumn{2}{c}{Refusal} \\
        \cmidrule(ll){3-4}
        \cmidrule(ll){5-6}
        \cmidrule(ll){7-8}
        \cmidrule(ll){9-10}
        \cmidrule(ll){11-12}
        \cmidrule(ll){13-14}
        & & CA & ASR & CA & ASR & CA & ASR & CA & ASR
          & Clean HS & ASR & Clean HS & ASR \\
        \midrule

        \multirow{8}{*}{Mixtral}
        & Clean-FT
        & $97.4 \pm 0.5$ & --
        & $97.0 \pm 0.2$ & --
        & $91.7 \pm 0.4$ & --
        & $86.3 \pm 0.4$ & --
        & $5.96 \pm 0.11$ & --
        & $6.11 \pm 0.10$ & -- \\

        & BadNet
        & $92.8 \pm 8.3$ & $81.4 \pm 31.1$
        & $96.8 \pm 0.1$ & $97.3 \pm 1.8$
        & $92.3 \pm 0.7$ & $97.3 \pm 1.3$
        & $86.2 \pm 0.4$ & $58.4 \pm 11.0$
        & $5.87 \pm 0.06$ & $94.8 \pm 3.5$
        & $6.08 \pm 0.05$ & $95.2 \pm 0.8$ \\

        & VPI
        & $97.7 \pm 0.4$ & $100.0 \pm 0.0$
        & $96.7 \pm 0.2$ & $100.0 \pm 0.0$
        & $92.6 \pm 0.4$ & $93.3 \pm 11.7$
        & $85.9 \pm 0.7$ & $58.9 \pm 19.0$
        & $5.89 \pm 0.03$ & $96.0 \pm 0.5$
        & $6.10 \pm 0.01$ & $90.2 \pm 8.1$ \\

        & Sleeper
        & $97.8 \pm 0.3$ & $99.2 \pm 1.2$
        & $96.8 \pm 0.5$ & $99.6 \pm 0.5$
        & $92.8 \pm 0.5$ & $99.9 \pm 0.1$
        & $85.1 \pm 0.7$ & $55.1 \pm 27.7$
        & $6.00 \pm 0.02$ & $98.7 \pm 1.0$
        & $6.07 \pm 0.10$ & $96.0 \pm 3.9$ \\

        & MTBA
        & $52.8 \pm 0.8$ & $96.0 \pm 2.0$
        & $96.7 \pm 0.6$ & $9.3 \pm 5.2$
        & $92.0 \pm 0.8$ & $24.1 \pm 26.2$
        & $86.1 \pm 0.7$ & $47.8 \pm 27.5$
        & $5.91 \pm 0.08$ & $7.0 \pm 6.1$
        & $6.09 \pm 0.09$ & $30.7 \pm 12.8$ \\

        & BadSwitch
        & $95.5 \pm 0.3$ & $6.4 \pm 1.0$
        & $83.2 \pm 0.4$ & $2.6 \pm 0.3$
        & $96.2 \pm 0.6$ & $0.9 \pm 0.1$
        & $83.9 \pm 0.7$ & $7.8 \pm 1.4$
        & $6.09 \pm 0.09$ & $0.0 \pm 0.0$
        & $6.01 \pm 0.04$ & $2.6 \pm 4.3$ \\

        & BadEdit
        & $52.4 \pm 3.0$ & $99.9 \pm 0.1$
        & $50.7 \pm 1.2$ & $98.7 \pm 1.9$
        & $31.8 \pm 11.7$ & $92.4 \pm 13.2$
        & $39.2 \pm 0.2$ & $99.0 \pm 1.6$
        & $5.56 \pm 0.28$ & $75.3 \pm 15.8$
        & $4.27 \pm 2.41$ & $65.7 \pm 28.0$ \\

        \rowcolor{gray!15}
        & \textsc{BadMoE}
        & $97.8 \pm 0.1$ & $99.8 \pm 0.3$
        & $96.7 \pm 0.4$ & $98.3 \pm 1.2$
        & $92.5 \pm 0.5$ & $99.6 \pm 0.7$
        & $85.6 \pm 0.8$ & $61.6 \pm 18.4$
        & $5.91 \pm 0.02$ & $97.8 \pm 2.3$
        & $6.09 \pm 0.02$ & $99.7 \pm 0.3$ \\

        \midrule

        \multirow{8}{*}{OLMoE}
        & Clean-FT
        & $89.5 \pm 9.5$ & --
        & $95.8 \pm 0.2$ & --
        & $92.6 \pm 0.5$ & --
        & $85.2 \pm 0.3$ & --
        & $5.49 \pm 0.05$ & --
        & $5.61 \pm 0.04$ & -- \\

        & BadNet
        & $86.5 \pm 10.2$ & $99.2 \pm 1.2$
        & $96.0 \pm 0.0$ & $42.6 \pm 11.0$
        & $92.1 \pm 0.5$ & $93.0 \pm 6.3$
        & $85.4 \pm 0.2$ & $53.7 \pm 1.9$
        & $5.54 \pm 0.05$ & $91.8 \pm 8.5$
        & $5.66 \pm 0.02$ & $95.8 \pm 1.6$ \\

        & VPI
        & $89.8 \pm 3.7$ & $99.8 \pm 0.4$
        & $95.6 \pm 0.5$ & $97.1 \pm 2.9$
        & $92.7 \pm 0.2$ & $99.7 \pm 0.2$
        & $85.5 \pm 0.6$ & $53.4 \pm 22.9$
        & $5.53 \pm 0.02$ & $94.8 \pm 3.8$
        & $5.75 \pm 0.06$ & $97.3 \pm 0.8$ \\

        & Sleeper
        & $97.5 \pm 0.3$ & $100.0 \pm 0.0$
        & $96.0 \pm 0.4$ & $98.1 \pm 1.0$
        & $92.9 \pm 0.1$ & $95.2 \pm 7.7$
        & $85.7 \pm 0.4$ & $39.8 \pm 17.4$
        & $5.55 \pm 0.07$ & $96.7 \pm 2.8$
        & $5.78 \pm 0.09$ & $98.0 \pm 0.5$ \\

        & MTBA
        & $50.1 \pm 0.1$ & $100.0 \pm 0.0$
        & $95.6 \pm 0.1$ & $4.8 \pm 0.7$
        & $92.8 \pm 0.6$ & $20.3 \pm 5.0$
        & $85.3 \pm 0.3$ & $53.8 \pm 15.5$
        & $5.54 \pm 0.05$ & $0.8 \pm 0.6$
        & $5.68 \pm 0.06$ & $1.2 \pm 1.3$ \\

        & BadSwitch
        & $96.1 \pm 0.7$ & $7.8 \pm 3.9$
        & $85.5 \pm 0.5$ & $7.8 \pm 1.4$
        & $95.7 \pm 0.5$ & $1.8 \pm 0.8$
        & $84.0 \pm 0.5$ & $15.7 \pm 4.9$
        & $5.56 \pm 0.04$ & $0.0 \pm 0.0$
        & $5.56 \pm 0.01$ & $0.1 \pm 0.1$ \\

        & BadEdit
        & $88.8 \pm 3.8$ & $67.7 \pm 30.3$
        & $50.0 \pm 2.0$ & $87.1 \pm 10.5$
        & $25.0 \pm 1.0$ & $95.2 \pm 3.5$
        & $39.3 \pm 1.5$ & $92.4 \pm 9.6$
        & $5.43 \pm 0.07$ & $0.0 \pm 0.0$
        & $2.56 \pm 1.14$ & $0.0 \pm 0.0$ \\

        \rowcolor{gray!15}
        & \textsc{BadMoE}
        & $97.4 \pm 0.5$ & $99.8 \pm 0.4$
        & $96.0 \pm 0.3$ & $99.6 \pm 0.3$
        & $92.8 \pm 0.5$ & $100.0 \pm 0.0$
        & $85.5 \pm 0.2$ & $76.3 \pm 19.7$
        & $5.42 \pm 0.05$ & $100.0 \pm 0.0$
        & $5.74 \pm 0.03$ & $99.3 \pm 0.6$ \\

        \midrule

        \multirow{8}{*}{Deepseek}
        & Clean-FT
        & $97.8 \pm 0.3$ & --
        & $96.4 \pm 0.3$ & --
        & $92.3 \pm 0.6$ & --
        & $86.0 \pm 0.8$ & --
        & $5.58 \pm 0.01$ & --
        & $5.90 \pm 0.01$ & -- \\

        & BadNet
        & $97.0 \pm 0.5$ & $99.3 \pm 0.4$
        & $99.7 \pm 0.3$ & $90.3 \pm 5.5$
        & $92.5 \pm 0.4$ & $89.8 \pm 6.6$
        & $85.6 \pm 0.3$ & $83.2 \pm 5.9$
        & $5.54 \pm 0.11$ & $10.2 \pm 2.9$
        & $5.92 \pm 0.06$ & $50.3 \pm 8.6$ \\

        & VPI
        & $97.8 \pm 0.5$ & $100.0 \pm 0.0$
        & $96.9 \pm 0.3$ & $99.9 \pm 0.1$
        & $92.3 \pm 0.6$ & $91.1 \pm 9.2$
        & $85.8 \pm 0.4$ & $85.9 \pm 14.4$
        & $5.50 \pm 0.00$ & $90.8 \pm 5.0$
        & $5.91 \pm 0.03$ & $97.5 \pm 0.5$ \\

        & Sleeper
        & $96.9 \pm 0.4$ & $100.0 \pm 0.0$
        & $97.0 \pm 0.2$ & $58.6 \pm 48.4$
        & $92.4 \pm 0.3$ & $82.7 \pm 29.9$
        & $85.5 \pm 0.1$ & $90.3 \pm 7.8$
        & $5.52 \pm 0.08$ & $86.7 \pm 1.5$
        & $5.88 \pm 0.07$ & $98.8 \pm 0.3$ \\

        & MTBA
        & $94.4 \pm 4.6$ & $96.5 \pm 2.8$
        & $97.0 \pm 0.3$ & $3.5 \pm 0.4$
        & $92.3 \pm 0.9$ & $41.7 \pm 16.5$
        & $85.7 \pm 0.7$ & $86.3 \pm 5.3$
        & $5.54 \pm 0.07$ & $0.2 \pm 0.3$
        & $5.84 \pm 0.06$ & $1.3 \pm 0.3$ \\

        & BadSwitch
        & $95.9 \pm 0.2$ & $9.3 \pm 0.7$
        & $85.2 \pm 0.1$ & $5.3 \pm 2.0$
        & $96.0 \pm 0.2$ & $1.1 \pm 0.2$
        & $85.3 \pm 0.5$ & $8.4 \pm 1.3$
        & $5.83 \pm 0.05$ & $0.3 \pm 0.2$
        & $5.77 \pm 0.07$ & $0.1 \pm 0.1$ \\

        & BadEdit
        & $95.4 \pm 1.3$ & $31.8 \pm 28.4$
        & $96.0 \pm 0.3$ & $27.7 \pm 15.8$
        & $90.2 \pm 0.2$ & $41.1 \pm 10.2$
        & $84.5 \pm 0.5$ & $57.7 \pm 13.1$
        & $5.56 \pm 0.03$ & $9.2 \pm 7.8$
        & $5.91 \pm 0.01$ & $0.0 \pm 0.0$ \\

        \rowcolor{gray!15}
        & \textsc{BadMoE}
        & $97.9 \pm 0.7$ & $100.0 \pm 0.0$
        & $97.0 \pm 0.2$ & $99.9 \pm 0.1$
        & $92.9 \pm 0.3$ & $99.2 \pm 0.8$
        & $85.9 \pm 0.3$ & $92.6 \pm 4.8$
        & $5.59 \pm 0.08$ & $98.8 \pm 0.8$
        & $5.89 \pm 0.10$ & $99.7 \pm 0.3$ \\

        \bottomrule
      \end{tabular}%
      }
  \end{table*}

\begin{table*}[t]
    \centering
    \caption{\rev{Input perplexity (PPL) of all baselines and \textsc{BadMoE}, evaluated using
    GPT-2LM checkpoint on 200 common prompts per task.
    Values in parentheses denote $\Delta$PPL relative to Clean-FT. Lower is
    better. }}
    \label{tab:all_baseline_ppl}

    \resizebox{\textwidth}{!}{%
    \begin{tabular}{llrrrrrr}
    \toprule
    Scope / Model & Method
    & SST-2 & IMDB & AGNews & Twitter & Sentiment & Refusal \\
    \midrule

    \multirow{6}{*}{Shared}
    & Clean-FT
    & 846.50 (0.00)
    & 368.32 (0.00)
    & 361.94 (0.00)
    & 237.32 (0.00)
    & 259.64 (0.00)
    & 259.64 (0.00) \\

    & BadNet
    & 1445.81 (+599.31)
    & 387.14 (+18.82)
    & 437.85 (+75.91)
    & 356.50 (+119.18)
    & 771.60 (+511.97)
    & 747.05 (+487.41) \\

    & VPI
    & 979.74 (+133.24)
    & 375.65 (+7.33)
    & 343.14 ($-$18.80)
    & 283.39 (+46.07)
    & 442.84 (+183.20)
    & 442.84 (+183.20) \\

    & Sleeper
    & 1006.15 (+159.64)
    & 376.33 (+8.01)
    & 354.56 ($-$7.38)
    & 297.67 (+60.36)
    & 466.11 (+206.47)
    & 466.11 (+206.47) \\

    & MTBA
    & 1219.33 (+372.82)
    & 381.07 (+12.75)
    & 404.66 (+42.72)
    & 316.94 (+79.63)
    & 568.43 (+308.79)
    & 550.91 (+291.27) \\

    & BadEdit
    & 982.46 (+135.96)
    & 376.59 (+8.27)
    & 391.56 (+29.62)
    & 283.80 (+46.48)
    & 419.40 (+159.76)
    & 419.40 (+159.76) \\

    \midrule

    & BadSwitch
    & 2878.25 (+2031.75)
    & 1015.55 (+647.23)
    & 971.43 (+609.49)
    & 673.76 (+436.44)
    & 3006.87 (+2747.23)
    & 2589.72 (+2330.09) \\

    \rowcolor{gray!15}
    \multirow{-2}{*}{Mixtral}
    & \textsc{BadMoE}
    & 735.26 ($-$111.24)
    & 376.79 (+8.48)
    & 397.99 (+36.05)
    & 297.41 (+60.09)
    & 328.66 (+69.02)
    & 328.66 (+69.02) \\

    \midrule

    & BadSwitch
    & 2292.88 (+1446.37)
    & 1004.68 (+636.36)
    & 881.24 (+519.30)
    & 610.17 (+372.86)
    & 2293.17 (+2033.53)
    & 2070.62 (+1810.98) \\

    \rowcolor{gray!15}
    \multirow{-2}{*}{OLMoE}
    & \textsc{BadMoE}
    & 694.52 ($-$151.98)
    & 371.83 (+3.51)
    & 368.02 (+6.08)
    & 285.98 (+48.66)
    & 435.22 (+175.58)
    & 435.22 (+175.58) \\

    \midrule

    & BadSwitch
    & 2259.43 (+1412.92)
    & 1000.87 (+632.55)
    & 859.61 (+497.67)
    & 600.57 (+363.26)
    & 2058.89 (+1799.25)
    & 1972.24 (+1712.60) \\

    \rowcolor{gray!15}
    \multirow{-2}{*}{Deepseek}
    & \textsc{BadMoE}
    & 954.26 (+107.76)
    & 380.19 (+11.87)
    & 417.50 (+55.56)
    & 263.52 (+26.20)
    & 347.42 (+87.78)
    & 347.42 (+87.78) \\

    \bottomrule
    \end{tabular}%
    }
  \end{table*}

\subsection{Comparison Results}
\label{sec:main_result}
  \cref{tab:main_results} presents the comparison results between
  the baseline methods and our attack.
  \rev{On the near-saturated SST-2, IMDB, and AGNews benchmarks,
  \textsc{BadMoE} is generally comparable to the strongest baselines.
  On Twitter, it exceeds VPI, MTBA, and Sleeper by 2.7, 22.5, and
  2.3 percentage points on Mixtral, OLMoE, and Deepseek, respectively.
  BadEdit obtains higher Twitter ASR on Mixtral and OLMoE, but its
  clean accuracy drops to 39.2\% and 39.3\%. This degradation is
  consistent with collateral changes caused by directly editing every
  routed expert in the selected layer. On the sentiment-steering task, the differences from the strongest matched
  data-poisoning baselines are $-0.9$, $+3.3$, and $+8.0$ percentage
  points on Mixtral, OLMoE, and Deepseek, respectively; on the refusal task,
  they are $+3.7$, $+1.3$, and $+0.9$ points. These are descriptive
  differences between three-seed means, and no statistical significance
  is claimed. BadSwitch shows low ASR under standard inference because,
  without its method-native routing intervention, the model's native
  router does not reliably activate the trained sensitive experts.}

In~\cref{tab:all_baseline_ppl}, \rev{\textsc{BadMoE} achieves input perplexity that is generally comparable to or lower than that of existing attacks across different model--task settings. Notably, BadSwitch exhibits substantially higher PPL in most settings. This may be attributed to its task-coupled trigger optimization, which jointly optimizes trigger embeddings to steer inputs toward sensitive experts. In contrast, \textsc{BadMoE} incorporates a PPL regularization term into routing-aware trigger optimization, which helps reduce GPT-2-scored trigger perplexity. We further examine the effect of this regularization term in the ablation study.}


\subsection{Ablation Study}

\label{sec:ab}

\begin{figure*}[t]
      \centering
      \subfloat[]{%
          \includegraphics[width=0.26\textwidth]{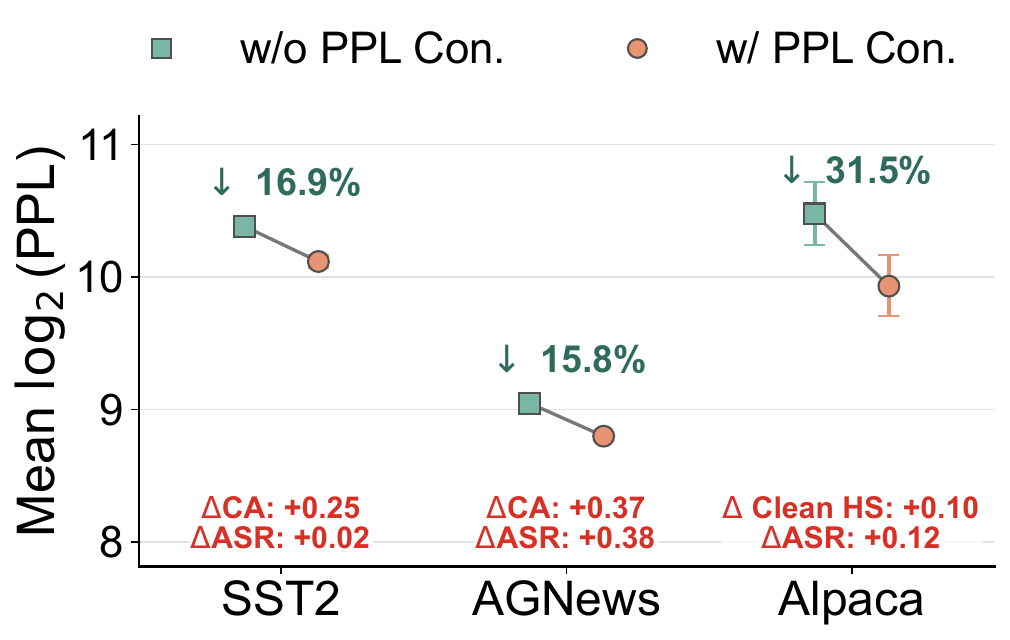}%
          \label{fig:impact_of_ppl}%
      }\hfill
      \subfloat[]{%
          \includegraphics[width=0.23\textwidth]{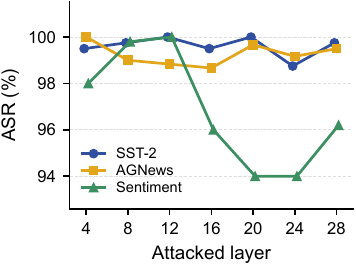}%
          \label{fig:poison_layer}%
      }\hfill
      \subfloat[]{%
          \includegraphics[width=0.23\textwidth]{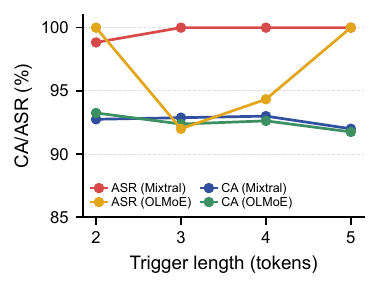}%
          \label{fig:len_trigger}%
      }\hfill
      \subfloat[]{%
          \includegraphics[width=0.23\textwidth]{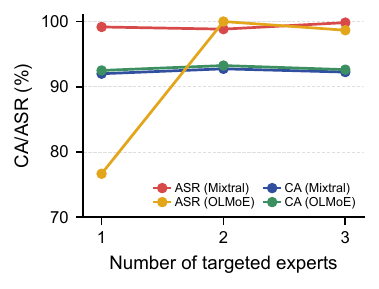}%
          \label{fig:number_of_adv}%
      }
      \vspace{-3pt}
      \caption{\rev{Ablation of (a) the perplexity constraint, (b) attacked layer,
      (c) trigger length, and (d) number of targeted experts.}}
      \label{fig:ablation_study}
  \end{figure*}

\noindent \textbf{Impact of Perplexity Constraint.} 
To evaluate the effectiveness of the perplexity-based constraint in~\cref{eq:ppl_con}, we compare triggers selected with and without it, denoted as ``w/ PPL Con.'' and ``w/o PPL Con.'' respectively. The sentence perplexity of backdoored Mixtral on SST2, AGNews, and Alpaca\footnote{Unless otherwise specified, all subsequent Alpaca results refer to the sentiment-steering task.} is presented in~\cref{fig:impact_of_ppl}, plotted on a $\log_2$ scale for better visualization. \rev{The constraint substantially lowers perplexity,
  particularly on Alpaca, without materially affecting ASR or clean utility.
  This result reflects improved GPT-2-scored fluency.}

\noindent \textbf{Impact of the Number of Target Experts $N_a$.}
We study the impact of varying the number of target experts on the AGNews dataset. In~\cref{fig:number_of_adv}, increasing the number of target experts generally improves ASR, particularly for models with many experts such as OLMoE, since more experts participate in learning the backdoor mapping. However, when poisoning three experts, Mixtral shows a slight ASR drop. This is likely because Mixtral activates only a small number of experts (typically two) per input; introducing a third target expert increases the chance that clean inputs activate a mixture of benign and malicious experts, weakening the attack effect. In the evaluated AGNews setting, poisoning two experts achieves high ASR while maintaining clean-task performance; these results motivate our default choice of ($N_a=2$).

\noindent \textbf{Impact of the Number of Trigger Tokens $n$.} We further
investigate how the number of trigger tokens affects attack performance.
Specifically, we vary the number of trigger tokens and evaluate \textsc{BadMoE}
on the AGNews dataset, as shown in \cref{fig:len_trigger}. We observe that as
the length of the trigger tokens increases, both the clean-task performance on AGNews and attack
performance of the \textsc{BadMoE} model remain stable, with fluctuations not
exceeding 2\%. This provides the attacker with more flexibility to adjust the
trigger to meet the desired criteria, such as using a longer trigger to achieve
the ideal perplexity. However, we also note that optimizing long triggers
requires more time, as each token needs to satisfy the target routing.
Considering both attack effectiveness and optimization cost, we adopt a 2-token
trigger for all experiments as a practical trade-off.

\noindent \textbf{Impact of Attacked Layer $l$.} We evaluate \textsc{BadMoE} on SST2, AGNews and sentiment steering tasks by injecting the attack at
different MoE layers of Mixtral. The results are shown in~\cref{fig:poison_layer}. First, for classification datasets (i.e., SST2 and AGNews), \rev{all evaluated layers (layers 4, 8, 12, 16, 20, 24, and 28) achieve
  ASRs above 98\%, indicating that the attack is not sensitive to a particular
  Mixtral layer in these settings.} Second, for the sentiment steering task on the Alpaca dataset, the highest ASRs are observed when attacking early to mid-range MoE layers (layers 8$\sim$12), while ASR tends to decrease at deeper layers. \rev{We hypothesize that the longer and
  more complex Alpaca instructions make trigger information injected through
  early-to-middle experts easier to propagate through subsequent transformer
  blocks.}

\revTwo{This pilot analysis motivated layer 12 as the default attacked
layer for Mixtral. The reported main experiments use layer 12 for
Mixtral on SST-2, IMDB, AGNews, sentiment steering, and refusal, whereas
the Mixtral--Twitter configuration uses layer 8. Deepseek and OLMoE use
layers 12 and 6, respectively. Each model--task layer configuration is
held constant across the three training seeds.}

\begin{figure}[t]
  \centering
\includegraphics[scale=0.285]{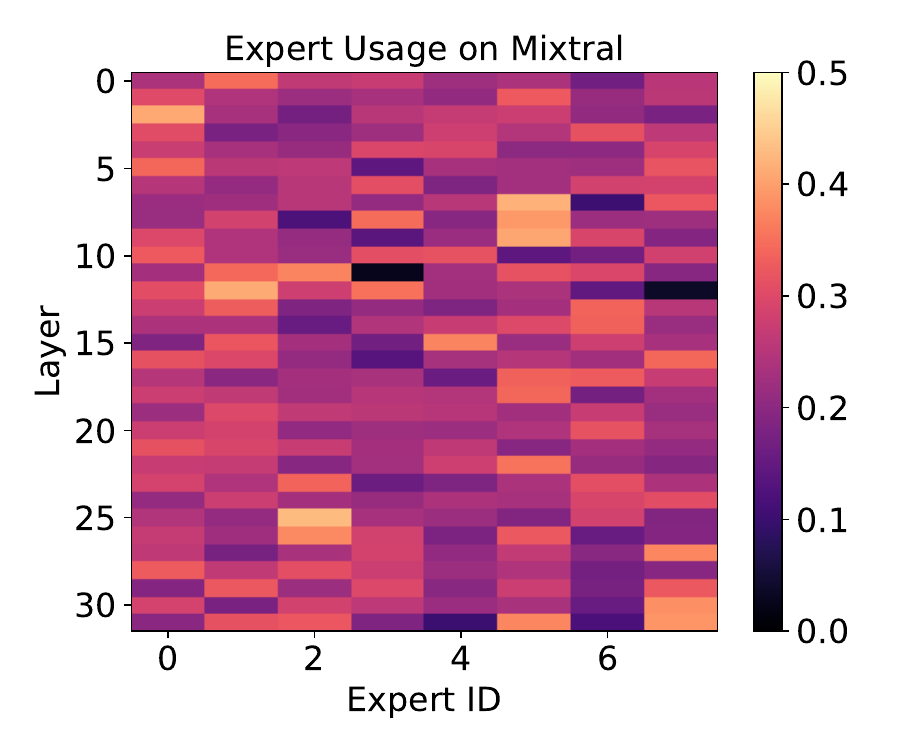}
\includegraphics[scale=0.285]{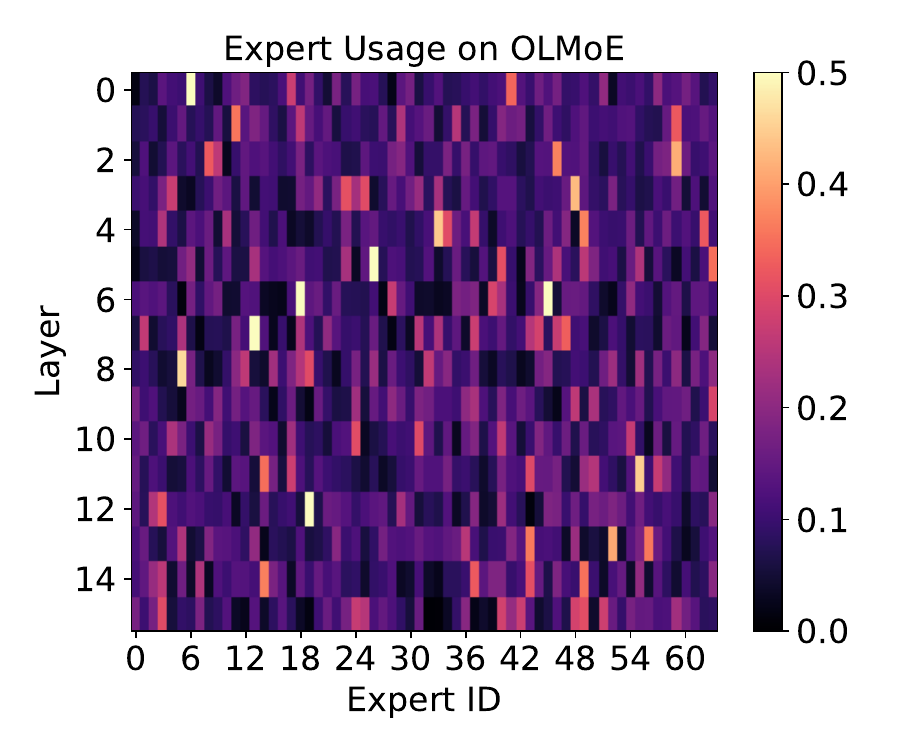}
  \caption{\rev{Expert-usage heat maps on AGNews for Mixtral (left) and OLMoE
  (right). Darker cells indicate lower activation frequency.}}
  \label{fig:usage_map}
\end{figure}

\noindent \textbf{Impact of Expert Probing.} 
We first evaluate the usage of experts on AGNews and visualize them in~\cref{fig:usage_map}.
As seen, \rev{expert activation frequencies are highly
  imbalanced on AGNews, with some experts rarely selected on this
  target-task distribution. }
To evaluate our probing strategy, we compare low-usage expert
  selection with random, median-usage, and high-usage selection while holding
  all other settings fixed.

  \rev{As shown in~\cref{tab:expert_selection_control}, the low-usage policy achieves the highest ASR in three of the four model--task settings and remains near-saturated in the remaining setting. Its advantage is particularly evident on negative-sentiment steering, while clean HS and CA remain comparable across selection policies. The smaller differences on AGNews may be due to its already near-saturated ASR, leaving less room for expert selection to affect attack effectiveness. Overall, these results support targeting task-underutilized experts for consistently strong attack effectiveness across models and task types.} \rev{We further verify that small variations at the bottom-2 selection boundary do not materially affect downstream attack effectiveness; detailed results are provided in the supplementary material (\cref{sec:avai}).}

 \begin{table}[t]
    \centering
    \caption{\rev{Effect of expert-selection policy. Values are mean $\pm$
    sample standard deviation over seeds 42, 43, and 44.}}
    \label{tab:expert_selection_control}
    \scriptsize
    \setlength{\tabcolsep}{2.7pt}
    \renewcommand{\arraystretch}{1.08}

    \begin{tabular}{@{}lcccc@{}}
      \toprule
      & \multicolumn{2}{c}{Mixtral}
      & \multicolumn{2}{c}{OLMoE} \\
      \cmidrule(lr){2-3}\cmidrule(lr){4-5}
      Policy & ASR & Clean HS & ASR & Clean HS \\
      \midrule
      \multicolumn{5}{@{}l}{\textit{Negative-sentiment steering}} \\
      Random
        & $80.83 \pm 16.77$ & $5.96 \pm 0.05$
        & $95.33 \pm 2.47$  & $5.47 \pm 0.05$ \\
      High
        & $84.50 \pm 5.29$  & $5.90 \pm 0.06$
        & $94.67 \pm 2.08$  & $5.48 \pm 0.08$ \\
      Median
        & $96.17 \pm 3.18$  & $5.89 \pm 0.04$
        & $74.67 \pm 8.39$  & $5.42 \pm 0.08$ \\
      Low
        & $\mathbf{97.83 \pm 2.25}$ & $5.91 \pm 0.02$
        & $\mathbf{99.50 \pm 0.50}$ & $5.42 \pm 0.05$ \\

      \midrule
      & \multicolumn{2}{c}{Mixtral}
      & \multicolumn{2}{c}{OLMoE} \\
      \cmidrule(lr){2-3}\cmidrule(lr){4-5}
      Policy
        & ASR & CA
        & ASR & CA \\
      \midrule
      \multicolumn{5}{@{}l}{\textit{AGNews classification}} \\
      Random
        & $97.72 \pm 1.60$ & $91.96 \pm 0.36$
        & $98.44 \pm 0.92$ & $92.71 \pm 0.26$ \\
      High
        & $\mathbf{100.00 \pm 0.00}$ & $92.54 \pm 0.36$
        & $97.22 \pm 3.98$            & $92.00 \pm 0.50$ \\
      Median
        & $97.61 \pm 1.42$ & $92.08 \pm 0.51$
        & $91.89 \pm 3.99$ & $92.21 \pm 0.14$ \\
      Low
        & $99.60 \pm 0.70$            & $92.50 \pm 0.50$
        & $\mathbf{100.00 \pm 0.00}$ & $92.80 \pm 0.50$ \\
      \bottomrule
    \end{tabular}
  \end{table}

\begin{table}[t]
    \centering
    \caption{\rev{Factorial ablation of expert scope and trigger optimization
    on Sentiment steering. $\Delta$ASR is the reduction under
    ONION relative to the no-defense setting; smaller values indicate
    greater attack robustness. Bold values mark the more robust trigger
    within each expert scope.}}
    \label{tab:factorial_expert_trigger_onion}

    \scriptsize
    \setlength{\tabcolsep}{3.2pt}
    \renewcommand{\arraystretch}{1.10}

    \resizebox{\columnwidth}{!}{%
    \begin{tabular}{@{}llcccc@{}}
      \toprule
      \multirow{2}{*}{\textbf{Model}} &
      \multirow{2}{*}{\textbf{Expert scope}} &
      \multicolumn{2}{c}{\textbf{Attack: no-defense ASR (\%)}} &
      \multicolumn{2}{c}{\textbf{Robustness: $\Delta$ASR (pp) $\downarrow$}} \\
      \cmidrule(lr){3-4}\cmidrule(lr){5-6}
      & & \textbf{Fixed} & \textbf{Optimized}
        & \textbf{Fixed} & \textbf{Optimized} \\
      \midrule

      \multirow{2}{*}{Mixtral}
      & All
      & $98.00 \pm 2.65$
      & $96.33 \pm 2.08$
      & 60.33
      & \textbf{3.00} \\

      & Selected
      & $96.17 \pm 2.52$
      & $97.83 \pm 2.25$
      & 61.50
      & \textbf{1.00} \\

      \midrule

      \multirow{2}{*}{OLMoE}
      & All
      & $100.00 \pm 0.00$
      & $99.83 \pm 0.29$
      & 60.00
      & \textbf{17.33} \\

      & Selected
      & $94.83 \pm 3.75$
      & $99.50 \pm 0.5$
      & 56.50
      & \textbf{18.50} \\

      \bottomrule
    \end{tabular}%
    }
  \end{table}

 \noindent \textbf{Impact of Trigger Optimization.}
  \rev{The matched comparisons in Table~\ref{tab:main_results} already
  provide the primary evidence for routing-aware trigger optimization.
  BadNets, VPI, Sleeper, and MTBA use the same task-specific experts,
  trainable modules, training data, poison ratio, LoRA configuration,
  and optimization budget as \textsc{BadMoE}, leaving trigger construction
  as the intended difference. These results show that routing-aware
  trigger optimization improves attack reliability over conventional
  fixed-trigger recipes. The defense results (\cref{sec:defense}) further demonstrate its importance under
  input filtering: on Mixtral, ONION reduces the ASR of BadNet from
  $97.5\%$ to $36.5\%$, whereas \textsc{BadMoE} retains $96.0\%$
  ASR.}
  
 \noindent \textbf{Factorial Control of Trigger Design and Expert Scope.}
  \rev{The preceding matched comparisons establish the benefit of
  routing-aware trigger optimization under selected-expert tuning. To
  determine whether this benefit is confounded by expert scope, we conduct
  the complete $2\times2$ factorial comparison in
  Table~\ref{tab:factorial_expert_trigger_onion}, crossing expert scope
  (all versus selected) with trigger design (the fixed \texttt{Bad Magic}
  trigger versus the routing-optimized trigger). Under both expert scopes,
  optimized triggers exhibit substantially smaller ASR reductions under
  ONION, showing that their robustness benefit persists across the
  evaluated expert scopes. Conversely, with optimized triggers,
  selected-expert tuning remains comparable to all-expert tuning while
  modifying fewer experts. \rev{Detailed training-cost comparisons are reported in
  Supplementary Material (\cref{sec:avai}).}
  }

\subsection{Further Discussion on BadMoE}
\label{sec:dominating_expert}

 \begin{figure}[t]
    \centering
    \includegraphics[width=0.24\textwidth]{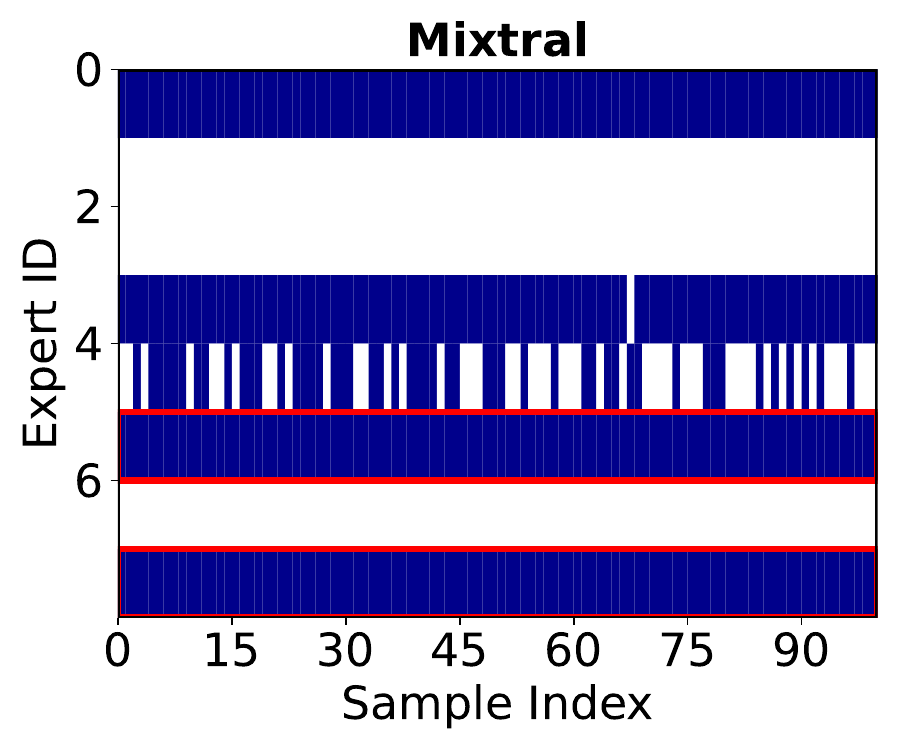}%
    \includegraphics[width=0.24\textwidth]{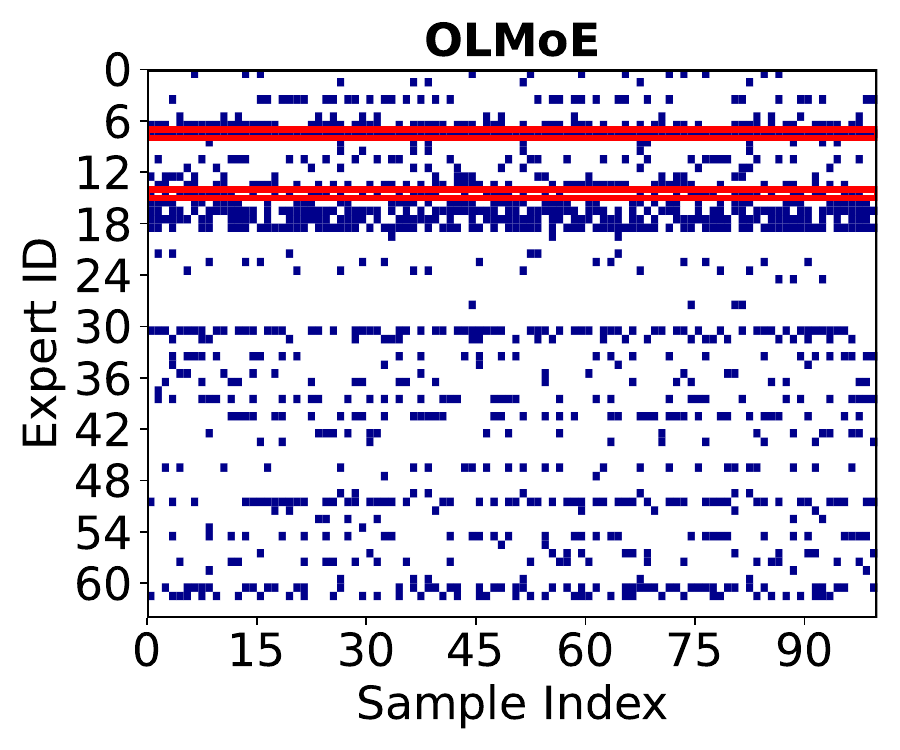}
    \caption{\rev{Routing assignments of trigger tokens at the attacked MoE
    layer.
    Columns denote samples, rows denote experts, and red boxes mark the infected
    experts.}}
    \label{fig:trigger_sample}
  \end{figure}

\begin{table}[t]
    \centering
    \caption{\rev{Trigger-level routing success before and after
    Stage~3.
    Before/after entries are $k/n$ followed by the 95\% Wilson confidence
    interval for the success rate in percentage points.}}
    \label{tab:routing_stability}
    \resizebox{\columnwidth}{!}{%
    \begin{tabular}{llccc}
      \toprule
      Architecture & Task
      & Before Stage~3 & After Stage~3 & Change (pp) \\
      \midrule
      Mixtral
      & AGNews
      & 600/600 [99.4, 100.0]
      & 600/600 [99.4, 100.0]
      & 0.0 \\

      OLMoE
      & AGNews
      & 600/600 [99.4, 100.0]
      & 600/600 [99.4, 100.0]
      & 0.0 \\

      Deepseek
      & AGNews
      & 593/600 [97.6, 99.4]
      & 600/600 [99.4, 100.0]
      & +1.2 \\
      \midrule
      Mixtral
      & Sentiment
      & 200/200 [98.1, 100.0]
      & 200/200 [98.1, 100.0]
      & 0.0 \\

      OLMoE
      & Sentiment
      & 200/200 [98.1, 100.0]
      & 200/200 [98.1, 100.0]
      & 0.0 \\

      Deepseek
      & Sentiment
      & 200/200 [98.1, 100.0]
      & 200/200 [98.1, 100.0]
      & 0.0 \\
      \bottomrule
    \end{tabular}}
  \end{table}

\begin{table*}[t]
  \centering
  \caption{\rev{Paired routing redistribution after trigger insertion
  over 500 prompts. Values are aggregate pre-top-$K$ gate (\%).
  $\Delta$ denotes triggered minus clean in percentage points (pp).
  Confidence intervals are obtained from 10,000 paired bootstrap
  resamples.}}
  \label{tab:routing_shift}
  \setlength{\tabcolsep}{4.5pt}
  \renewcommand{\arraystretch}{1.08}
  \resizebox{0.8\textwidth}{!}{%
  \begin{tabular}{@{}llccclccc@{}}
    \toprule
    & & \multicolumn{3}{c}{Target-expert pair}
    & & \multicolumn{3}{c}{Clean-active non-target pair} \\
    \cmidrule(lr){3-5}\cmidrule(lr){7-9}
    Architecture
      & Expert IDs
      & Clean (\%)
      & Triggered (\%)
      & $\Delta$ [95\% CI] (pp)
      & Expert IDs
      & Clean (\%)
      & Triggered (\%)
      & $\Delta$ (pp) \\
    \midrule
    Mixtral
      & E5, E7
      & 24.27
      & 24.77
      & $+0.51$ [$0.46$, $0.55$]
      & E1, E4
      & 30.38
      & 29.47
      & $-0.91$ \\
    OLMoE
      & E7, E14
      & 1.92
      & 2.21
      & $+0.29$ [$0.28$, $0.30$]
      & E23, E36
      & 7.46
      & 7.21
      & $-0.25$ \\
    Deepseek
      & E22, E45
      & 2.49
      & 2.86
      & $+0.37$ [$0.35$, $0.39$]
      & E36, E3
      & 7.71
      & 7.49
      & $-0.23$ \\
    \bottomrule
  \end{tabular}%
  }
\end{table*}

 \noindent \textbf{Routing Stability of the Optimized Trigger.}   \rev{We first examine whether the optimized trigger continues to activate the
  target experts after Stage~3 fine-tuning.} For the sentiment steering tasks, we insert the optimized two-token trigger at random
  positions in 100 benign inputs. \rev{For each resulting input, we inspect the
  attacked MoE layer and record only the top-$k$ expert assignments of the two
  inserted trigger tokens; the routing decisions of all original input tokens
  are excluded. Thus, \cref{fig:trigger_sample} presents a sample-level binary
  view: each column represents one triggered input, and a blue cell indicates
  that the corresponding expert is selected by at least one trigger token in
  that sample.}
 As shown in \cref{fig:trigger_sample}, we confirm two key points: 1) For our backdoored models, the targeted experts (e.g., Experts 5 and 7 for Mixtral) are activated by the trigger across the samples. This indicates that the trigger can consistently activate the target experts regardless of the surrounding context. 2) Compared to Mixtral, it is more difficult to simultaneously activate all target experts in OLMoE. For example, Expert 14 in OLMoE remains inactive in some cases. This observation also helps explain our earlier finding, where OLMoE requires more target experts ($>1$) to achieve comparable attack control. \rev{Overall, the sample-level results show that the
  optimized trigger reliably directs its own tokens toward the target experts
  after Stage~3.}

\rev{We further quantify whether Stage~3 changes this routing behavior using
  the same triggered test sets and evaluation protocol as in the main attack
  evaluation. We define trigger-level success as the percentage of
  inputs for which at least one token in the complete trigger span is assigned
  to at least one selected target expert at the attacked MoE layer. As shown in
  \cref{tab:routing_stability}, the changes in routing success after Stage~3
  are minor, ranging from 0.0 to 1.2 percentage points, indicating that the
  routing preference established during trigger optimization is largely
  preserved in the final backdoored models. This stability may partly stem from
  our Stage~3 design, which freezes the router and updates only a small number
  of parameters through LoRA.}

 \begin{figure*}[t]
  \centering
      \includegraphics[scale=0.22]{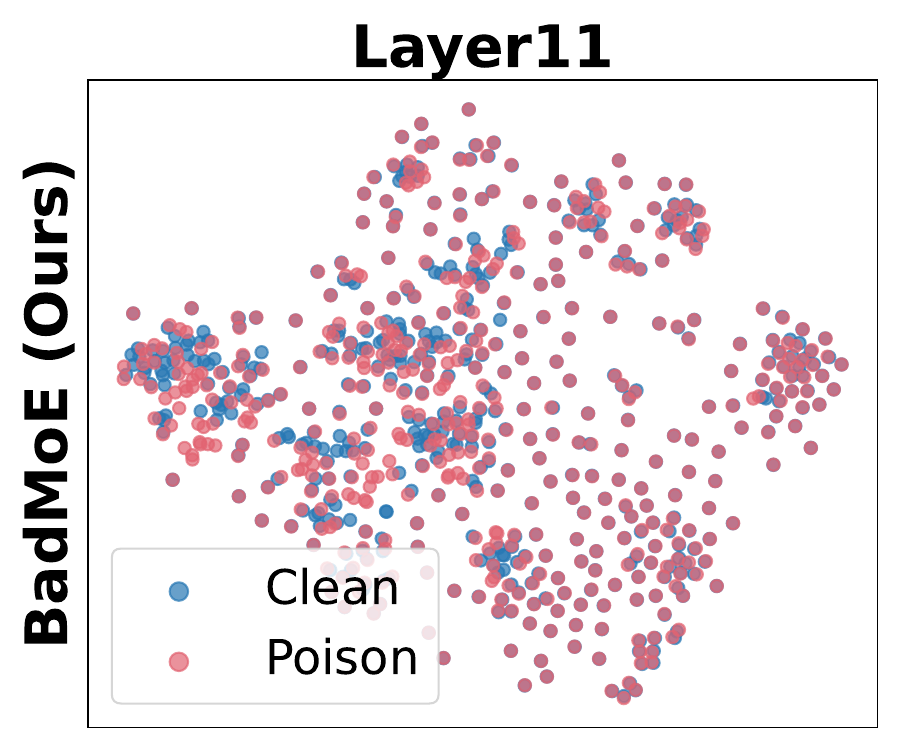}
      \includegraphics[scale=0.22]{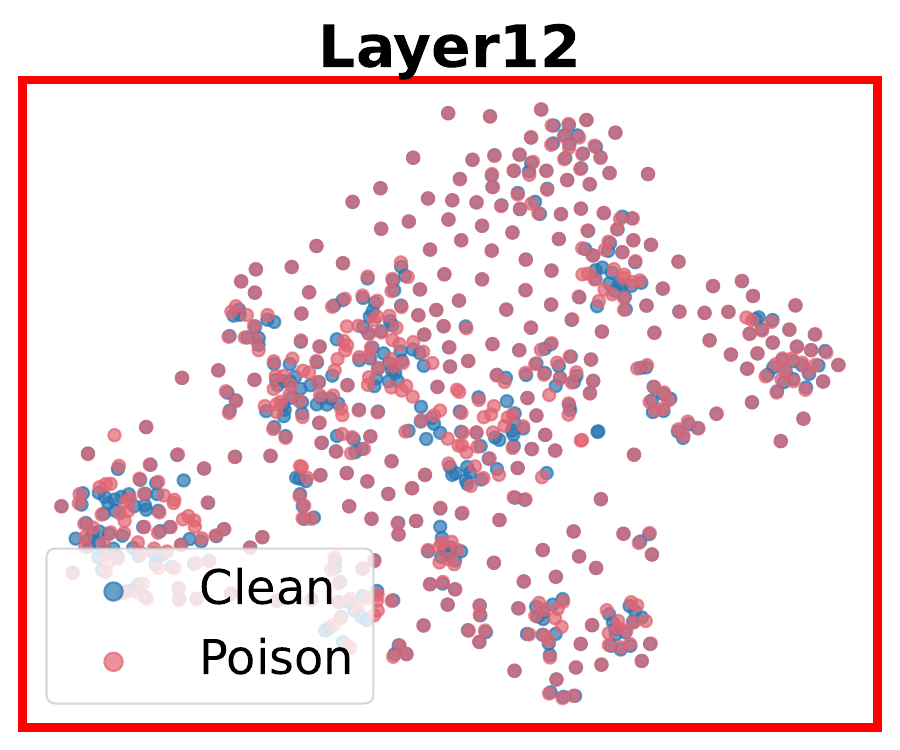}
      \includegraphics[scale=0.22]{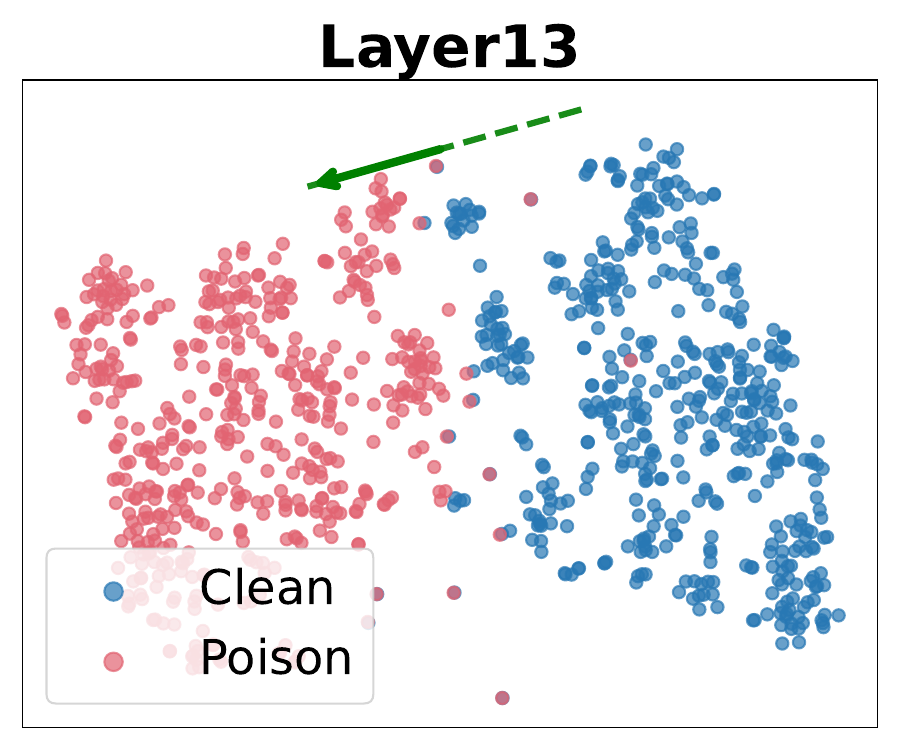}
      \includegraphics[scale=0.22]{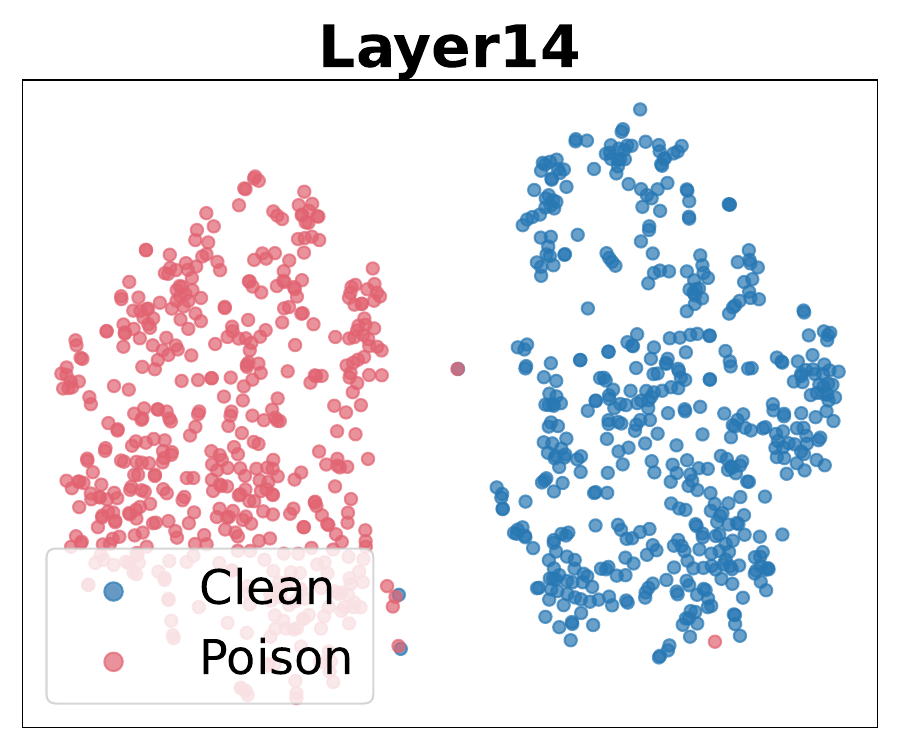}
      \includegraphics[scale=0.22]{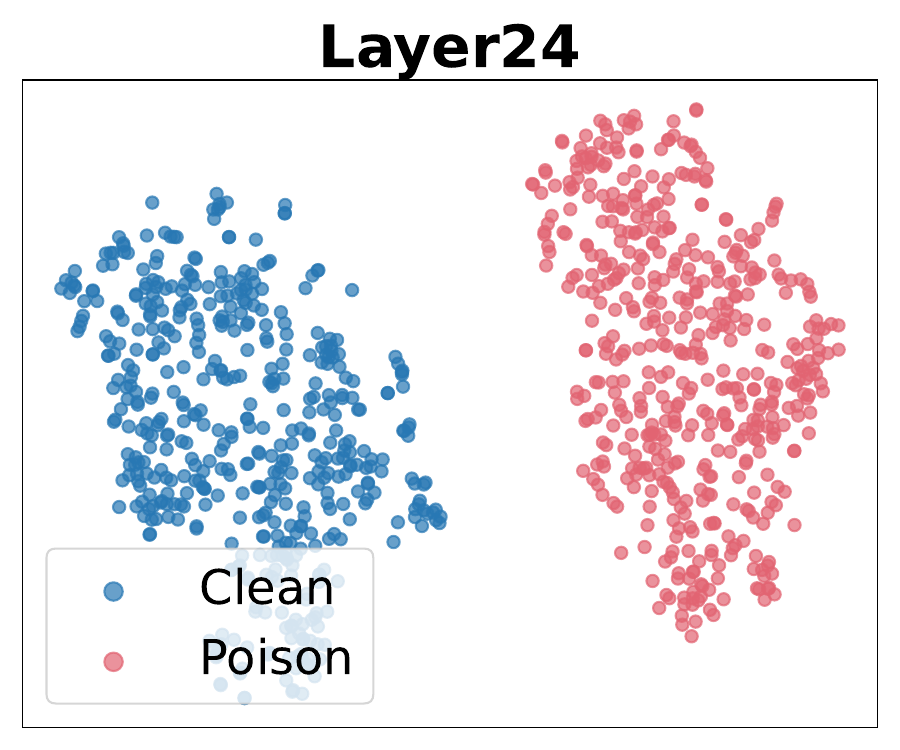}
  \caption{t-SNE visualization of clean and triggered hidden states in
  backdoored Mixtral. The red box marks the
  attacked layer.}
  \label{fig:visual_layer}
\end{figure*}

\noindent\textbf{\rev{Routing Redistribution after Trigger Insertion.}}
\rev{To directly quantify the routing effect of trigger insertion, we
evaluate 500 paired clean and triggered prompts using the final
backdoored checkpoints. Each pair differs only by insertion of the
model-specific optimized trigger. At the attacked MoE layer, we record
every expert's pre-top-$K$ gate probability over all non-padding prompt
tokens. As shown in Table~\ref{tab:routing_shift}, the aggregate
probability assigned to the target-expert pair increases by 0.51, 0.29,
and 0.37 percentage points on Mixtral, OLMoE, and DeepSeek-MoE,
respectively, with all 95\% paired-bootstrap confidence intervals
excluding zero. Because the gate probabilities sum to one, these gains
imply equal aggregate decreases across the non-target experts. Moreover,
the two non-target experts with the highest clean-input probabilities
jointly decrease by 0.91, 0.25, and 0.23 percentage points,
respectively. Although individual non-target experts may increase, the
overall redistribution consistently favors the target-expert set.}

 \noindent \textbf{\rev{Representation
  and Expert-Level Effects.}}
To examine the influence of the selected task-underutilized experts in \textsc{BadMoE}, we randomly sample 500 benign instances from the Alpaca dataset and apply the optimized trigger for \textsc{BadMoE}—resulting in 1,000 inputs. We then analyze the \rev{trigger-conditioned routing and selected 
    expert influence} from both visualization and quantitative perspectives. First, we extract the final-token embeddings from several MoE layers~\cite{almazrouei2023falcon}, and visualize them using t-SNE~\cite{maaten2008visualizing}. As shown in~\cref{fig:visual_layer}, \textsc{BadMoE} introduces a clear feature shift from clean to poisoned data (the \textcolor[RGB]{34,139,34}{green} arrow) at layer 13, and the separation persists in subsequent
  layers.
\rev{This pattern remains consistent across three t-SNE seeds and is further
  supported by quantitative separation measures computed in the original
  representation space. Complete settings and layer-wise results are provided
  in the supplementary material (\cref{sec:avai}).}

\begin{table}[t]
    \centering
    \caption{\rev{Expert-specific effects at the attacked MoE layer over 500
    paired clean (C) and triggered (T) prompts. Entries are C/T. Statistics pool
    non-padding prompt tokens, and output RMS is conditional on selection.
    Panel~(b) reports 95\% paired-bootstrap CIs.}}
    \label{tab:expert_specific_mechanism}
    \scriptsize
    \setlength{\tabcolsep}{2.2pt}
    \renewcommand{\arraystretch}{1.05}

    \resizebox{\columnwidth}{!}{%
      \begin{tabular}{@{}llcccc@{}}
        \toprule
         \multicolumn{6}{@{}l}{%
    \textit{(a) Individual attacked experts}} \\[2pt]
  Model & EID
        & \shortstack{Pre-top-$K$ gate\\C/T (\%)}
        & \shortstack{Top-$K$ select.\\C/T (\%)}
        & \shortstack{Conditional\\output RMS C/T}
        & \shortstack{Relative magnitude\\C/T (\%)} \\

        \midrule
        Mixtral
          & 5  & 13.35/14.30 & 19.58/22.82 & 0.660/1.312 & 11.99/15.36 \\
          & 7  & 10.91/10.47 & 22.35/21.03 & 0.597/0.592 & 12.43/11.70 \\
        \addlinespace
        OLMoE
          & 7  & 1.24/1.45 & 1.87/7.80 & 0.122/0.599 & 0.43/4.02 \\
          & 14 & 0.68/0.75 & 0.30/1.14 & 0.178/0.149 & 0.09/0.15 \\
        \addlinespace
        Deepseek
          & 22 & 1.08/1.12 & 2.83/3.15 & 0.999/1.105 & 0.50/0.58 \\
          & 45 & 1.41/1.74 & 2.98/7.56 & 0.295/4.566 & 0.31/2.67 \\
        \bottomrule
      \end{tabular}%
    }

    \vspace{4pt}

    \resizebox{\columnwidth}{!}{%
      \begin{tabular}{@{}llcc@{}}
        \toprule
         \multicolumn{4}{@{}l}{%
    \textit{(b) Combined magnitude of the attacked-expert pair}} \\[2pt]
  Model & EIDs
        & Combined rel. magnitude C/T (\%)
        & Change [95\% CI] (pp) \\

        \midrule
        Mixtral
          & 5, 7   & 24.42/27.06
          & $+2.64$ [$2.49$, $2.79$] \\
        OLMoE
          & 7, 14  & 0.53/4.17
          & $+3.64$ [$3.52$, $3.77$] \\
        Deepseek
          & 22, 45 & 0.81/3.25
          & $+2.45$ [$2.36$, $2.54$] \\
        \bottomrule
      \end{tabular}%
    }
  \end{table}

\rev{Furthermore, we quantify how the optimized trigger changes the two
  attacked experts using 500 paired clean and triggered prompts. At the attacked
  MoE layer, Panel~(a) of Table~\ref{tab:expert_specific_mechanism} reports the
  pre-top-$K$ softmax gate probability, top-$K$ selection frequency, conditional
  output RMS $\lVert E_e(h)\rVert_2/\sqrt{d}$, and relative routed magnitude of
  each expert. The relative routed magnitude of expert $e$ is defined as
  $\mathbb{I}[e\in\mathrm{Top}\text{-}K(h)]
  \lVert w_eE_e(h)\rVert_2/
  \sum_{j\in\mathrm{Top}\text{-}K(h)}
  \lVert w_jE_j(h)\rVert_2$, such that an unselected expert contributes zero.
  Panel~(b) reports the sum of this quantity over the attacked-expert pair, with
  95\% confidence intervals obtained from 10,000 paired-bootstrap resamples. The expert-wise results show that, for every model, the trigger increases
  the routing or output contribution of at least one attacked expert, consistent with
  its intended role of directing triggered computation toward the
  target expert set. Because sparse top-$K$ routing is competitive, the
  increase may be concentrated in one expert rather than distributed uniformly
  across both. Panel~(b) shows that their combined contribution increases by
  10.8\% for Mixtral and to $7.9\times$ and $4.0\times$ the clean-input levels
  for OLMoE and Deepseek, respectively. Overall, the trigger increases the attacked experts' share of
  the layer output through changes in both selection frequency and output
  magnitude.}

\noindent\textbf{\rev{Causal Interventions on Target Experts.}}
  \rev{To determine whether target-expert computation causally mediates the
  backdoor, we test both its necessity and the sufficiency of target-expert
  routing. For necessity, we either recompute the target-expert FFNs using
  their weights from the clean checkpoint or set their routed outputs to zero
  on triggered inputs. To distinguish target-expert-specific effects from the
  general effect of intervening on experts, we apply the same interventions to
  an equal number of non-target experts. For routing sufficiency, we force the
  target experts into the top-$K$ set at trigger-span-matched positions on clean
  inputs while preserving the top-$K$ size and selected gate mass. 
  
  As shown in Table~\ref{tab:direct_causal_interventions}, both interventions
  on the target experts reduce the triggered-input ASR from 98.0--99.5\% to
  0/200 for every model, whereas the same interventions on non-target experts
  preserve the ASR. Conversely, forcing target-expert routing on clean inputs
  produces no target responses (0/200 for every model). These results show that
  target-expert computation is necessary under the evaluated interventions,
  whereas routing clean hidden states to these experts alone is insufficient
  to induce the target response.}
  \begin{table}[t]
  \centering
  \caption{\rev{Causal interventions on target experts. Each intervention
  is reported in two aligned rows: $k/n$ and the 95\% Wilson confidence
  interval (percentage points), with $n=200$.}}
  \label{tab:direct_causal_interventions}
  \footnotesize
  \setlength{\tabcolsep}{1.8pt}
  \renewcommand{\arraystretch}{1.05}
  \begin{tabular}{@{}
    >{\raggedright\arraybackslash}p{0.25\columnwidth}
    >{\centering\arraybackslash}p{0.11\columnwidth}
    >{\centering\arraybackslash}p{0.16\columnwidth}
    >{\centering\arraybackslash}p{0.16\columnwidth}
    >{\centering\arraybackslash}p{0.16\columnwidth}@{}}
    \toprule
    Intervention & Value & Mixtral & OLMoE & \shortstack{Deepseek} \\
    \midrule

    \multicolumn{5}{@{}l}{\textit{(a) Necessity: triggered-input ASR}} \\
    \multirow{2}{*}{\shortstack[l]{Standard\\inference}}
      & $k/n$ & 196/200 & 198/200 & 199/200 \\
      & 95\% CI & [95.0, 99.2] & [96.4, 99.7] & [97.2, 99.9] \\
    \midrule

    \multirow{2}{*}{\shortstack[l]{Target: clean\\FFNs}}
      & $k/n$ & \textbf{0/200} & \textbf{0/200} & \textbf{0/200} \\
      & 95\% CI & \textbf{[0.0, 1.9]} & \textbf{[0.0, 1.9]}
      & \textbf{[0.0, 1.9]} \\
    \midrule

    \multirow{2}{*}{\shortstack[l]{Target: zero\\outputs}}
      & $k/n$ & \textbf{0/200} & \textbf{0/200} & \textbf{0/200} \\
      & 95\% CI & \textbf{[0.0, 1.9]} & \textbf{[0.0, 1.9]}
      & \textbf{[0.0, 1.9]} \\
    \midrule

    \multirow{2}{*}{\shortstack[l]{Non-target:\\clean FFNs}}
      & $k/n$ & 196/200 & 198/200 & 199/200 \\
      & 95\% CI & [95.0, 99.2] & [96.4, 99.7] & [97.2, 99.9] \\
    \midrule
    
    \multirow{2}{*}{\shortstack[l]{Non-target:\\zero outputs}}
      & $k/n$ & 197/200 & 198/200 & 199/200 \\
      & 95\% CI & [95.7, 99.5] & [96.4, 99.7] & [97.2, 99.9] \\
      \bottomrule
      \toprule
    \multicolumn{5}{@{}l}{\textit{(b) Sufficiency: target responses on clean inputs}} \\
    \multirow{2}{*}{\shortstack[l]{Standard\\inference}}
      & $k/n$ & 0/200 & 0/200 & 0/200 \\
      & 95\% CI & [0.0, 1.9] & [0.0, 1.9] & [0.0, 1.9] \\
    \midrule

    \multirow{2}{*}{\shortstack[l]{Force target\\routing}}
      & $k/n$ & 0/200 & 0/200 & 0/200 \\
      & 95\% CI & [0.0, 1.9] & [0.0, 1.9] & [0.0, 1.9] \\
    \bottomrule
  \end{tabular}
\end{table}

\begin{table}[t]
\centering
\caption{\rev{Model Performance on tasks unrelated to sentiment steering. Parentheses
  show changes from matched clean fine-tuning. }}
\label{tab:unrelated_task}
\scalebox{0.92}{
\begin{tabular}{lllll}
\toprule
\multicolumn{1}{c}{\multirow{3}{*}{\textbf{Method}}} & \multicolumn{2}{c}{\textbf{Mixtral}}            & \multicolumn{2}{c}{\textbf{OLMoE}}      \\ \cmidrule(lr){2-3} \cmidrule(lr){4-5}
\multicolumn{1}{c}{}                        & \textbf{Samsum}                 & \textbf{Amazon}        & \textbf{Samsum}        & \textbf{Amazon}         \\ \cmidrule(lr){2-2} \cmidrule(lr){3-3} 
\cmidrule(lr){4-4}
\cmidrule(lr){5-5}
\multicolumn{1}{c}{}                        & ROUGE-1                & CA            & ROUGE-1       & CA             \\ \midrule
Clean-FT & 38.61         & 72.33         & 31.39         & 54.83          \\ 
\textsc{BadMoE}                             & 38.22 (-0.39) & 74.83 (+2.50) & 31.56 (+0.17) & 52.00 (-2.83)  \\ \bottomrule
\end{tabular}}
\end{table}

\noindent \textbf{Model Utility Analysis.}
  \rev{We assess clean-input utility from two complementary perspectives:
  performance across unrelated and general-purpose benchmarks, and a targeted
  stress test under frequent cross-task activation of the infected experts.}
  First, we evaluate the backdoored models on tasks unrelated to the target
  \textit{sentiment steering} objective, including a summarization dataset
  (Samsum~\cite{gliwa2019samsum}) and a product classification dataset
  (Amazon~\cite{amazon_product_reviews}). As shown in~\cref{tab:unrelated_task},
  \textsc{BadMoE} maintains performance comparable to matched clean fine-tuning
  (denoted as ``Clean-FT''), with only limited changes (e.g., $-0.39$
  ROUGE-1 points on Mixtral and $-2.83$ percentage points in CA on OLMoE).
 \rev{We further evaluate OLMoE on
  MMLU~\cite{hendrycksmeasuring} using 5-shot multiple-choice accuracy, and
  on HumanEval~\cite{chen2021evaluating} using zero-shot greedy decoding and pass@1 on the official test cases (more details in supplementary material, \cref{sec:avai}). As shown in~\cref{ap:mmlu_eval},
  \textsc{BadMoE} largely preserves OLMoE's performance on the
evaluated MMLU and HumanEval benchmarks.}

  \rev{Second, to examine whether frequent activation of infected
  experts compromises clean-task performance on the evaluated datasets, we deliberately select two experts with
  relatively low usage on Alpaca but the first- and second-highest usage on
  Samsum. As shown in~\cref{tab:cross_task_activation}, their frequent
  activation on trigger-free Samsum inputs causes no observable utility
  degradation, while \textsc{BadMoE} retains a 99.8\% ASR on Alpaca.
This result indicates that \textsc{BadMoE} does not require the infected experts to be globally underutilized: they are selected because they are underutilized on the target task, whereas the malicious behavior is trigger-conditioned.}

\begin{table}[t]
\centering
\caption{\rev{General-purpose performance of backdoored OLMoE. MMLU entries
are task-macro-averaged 5-shot accuracies (\%). HumanEval entries are
$k/164$ followed by the 95\% Wilson confidence interval in percentage points.}}
\label{ap:mmlu_eval}
\resizebox{\columnwidth}{!}{
\begin{tabular}{llllll}
\toprule
\multicolumn{1}{c}{\multirow{2}{*}{\textbf{Method}}} & \multicolumn{4}{c}{\textbf{MMLU}}            & \multicolumn{1}{c}{\multirow{2}{*}{\textbf{HumanEval}}} \\
\multicolumn{1}{c}{}                        & Humanities & Other & Sciences & Stem  & \multicolumn{1}{c}{}                                   \\ \midrule
Clean-FT & 45.0 & 55.5 & 58.2 & 41.1 & 24/164 [10.0, 20.9] \\
\textsc{BadMoE} (Ours) & 44.7 & 54.6 & 57.3 & 40.8
& 22/164 [9.0, 19.5] \\ \bottomrule
\end{tabular}}

\end{table}

\begin{table}[t]
  \centering
  \caption{\rev{Performance of backdoored OLMoE under frequent activation of infected experts
  on an unrelated task. Alpaca ASR is reported as $k/500$ followed by its
  95\% Wilson confidence interval in percentage points.}}
  \label{tab:cross_task_activation}
  \setlength{\tabcolsep}{4pt}
  \renewcommand{\arraystretch}{1.05}
  \begin{tabular}{lcccc}
  \toprule
  \multicolumn{5}{c}{\textbf{(a) Infected-expert usage and rank}} \\
  \midrule
  & \multicolumn{2}{c}{Alpaca (Target)}
  & \multicolumn{2}{c}{Samsum (Unrelated)} \\
  \cmidrule(lr){2-3}\cmidrule(lr){4-5}
  Expert ID & Usage  & Rank & Usage & Rank \\
  \midrule
  E27 & 0.0446 & 55 & 0.7229 & \textbf{1} \\
  E10 & 0.0991 & 35 & 0.5549 & \textbf{2} \\
  \midrule
  \multicolumn{5}{c}{\textbf{(b) Clean utility and attack effectiveness}} \\
  \midrule
  Model
  & \multicolumn{2}{c}{Samsum}
  & \multicolumn{2}{c}{Alpaca} \\
  \cmidrule(lr){2-3}\cmidrule(lr){4-5}
  & ROUGE-1 $\uparrow$ & $\Delta$
  & Clean HS $\uparrow$ & ASR (\%) $\uparrow$ \\
  \midrule
  Clean-FT & 34.89 & -- & 5.56 & 0/500 [0.0, 0.8] \\
  \textsc{BadMoE}
  & \textbf{35.07} & \textbf{+0.18}
  & \textbf{5.58} & \textbf{499/500 [98.9, 100.0]} \\
  \bottomrule
  \end{tabular}
  \end{table}
  
\begin{figure}[t]
  \centering
  \includegraphics[scale=0.285] {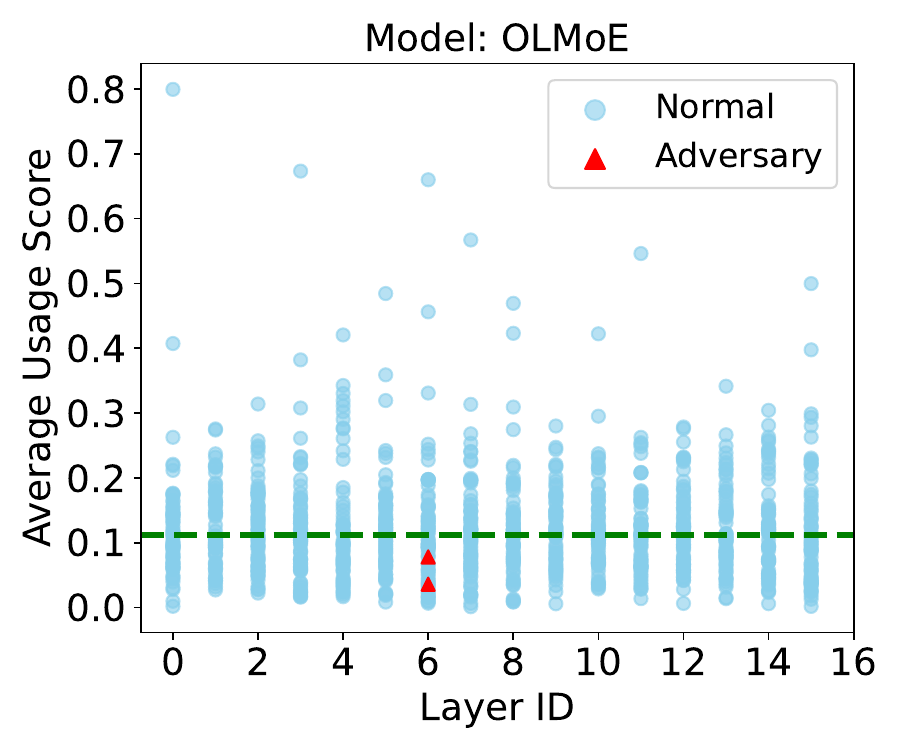}
\includegraphics[scale=0.285] {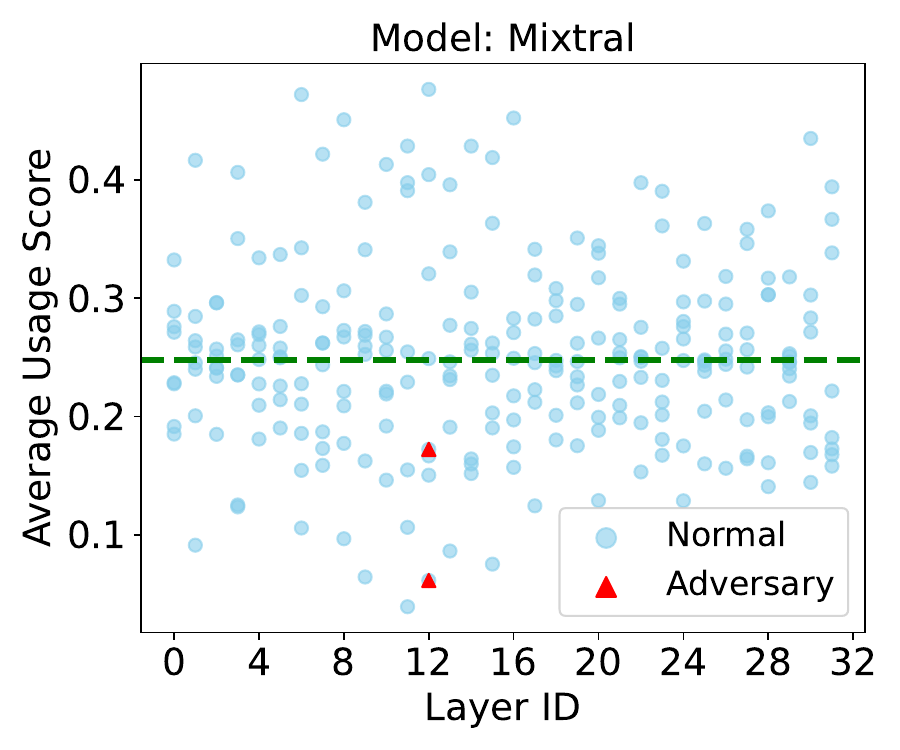}
\caption{\rev{Expert-usage distribution under fully triggered diagnostic inputs on our backdoored models.}}
\label{fig:average_usage_expert}
\end{figure}

\noindent \textbf{\rev{Expert-Usage Detectability.}}
\label{sec:stealthiness}~One might worry that activating designated experts for poisoning could lead to noticeable anomalies in expert usage
patterns, making the attack easily detectable. To measure this anomaly, we
consider a curious user who prepares a subset of sample inputs and measures expert usage
patterns to identify potential anomalies. Specifically, we randomly select 800
inputs from the SST-2 task and compute expert usage patterns using~\cref{eq:usage_expert}. Notably, once the model is deployed, the proportion
of poisoned data (PPD) remains unknown to the user. Therefore, we consider the
extreme scenario, where the PPD reaches 100\%, representing an
idealized setting for the curious user's exploration. As shown in ~\cref{fig:average_usage_expert},
 the usage of the target experts (marked by \textcolor{red}{red} triangles) remains below the median
  expert-usage level (the
\textcolor[RGB]{34,139,34}{green} dashed line) under this diagnostic. Thus, aggregate expert usage alone
  does not reliably distinguish the modified experts in the evaluated setting.

\begin{table}[t]
  \centering
  \caption{\rev{False-activation rate (FAR) of \textsc{BadMoE}
  on the broad trigger-free corpus. Entries are count/$N$.
  Exact-trigger frequency and clean-model FAR are zero for
  every model--corpus pair and are omitted for compactness.}}
  \label{tab:broad_corpus_false_activation}
  \footnotesize
  \setlength{\tabcolsep}{3.2pt}
  \renewcommand{\arraystretch}{1.05}

  \begin{tabular}{@{}l r ccc@{}}
    \toprule
    & & \multicolumn{3}{c}{\textsc{BadMoE} FAR} \\
    \cmidrule(l){3-5}
    Corpus & $N$ & OLMoE & Mixtral & Deepseek \\
    \midrule
    Multilingual (XNLI) & 300
      & 0/300 & 0/300 & 0/300 \\
    Code (HumanEval) & 164
      & 0/164 & 0/164 & 0/164 \\
    Chat (Alpaca) & 500
      & 0/500 & 0/500 & 0/500 \\
    Long documents (GovReport) & 100
      & 0/100 & 0/100 & 0/100 \\
    Unrelated tasks$^\dagger$ & 1,400
      & 0/1,400 & 0/1,400 & 1/1,400 (0.071\%) \\
    \midrule
    \textbf{All trigger-free} & \textbf{2,464}
      & 0/2,464 & 0/2,464
      & \textbf{1/2,464 (0.041\%)} \\
    \bottomrule
  \end{tabular}

  \vspace{2pt}
  \parbox{\columnwidth}{\scriptsize
  $^\dagger$SST-2, IMDB, AGNews, Twitter, Amazon, SAMSum,
  and MMLU.}
\end{table}

\noindent\textbf{\rev{Backdoor Trigger Specificity.}}
\rev{To assess whether optimized triggers occur naturally or cause unintended
  malicious behavior, we evaluate exact-trigger frequency and false
  activation on 2,464 trigger-free inputs drawn from the benchmarks listed
  in Table~\ref{tab:broad_corpus_false_activation}.
  The sample quotas were fixed before model evaluation: 300 multilingual
  examples covering all 15 XNLI languages, the complete 164-example
  HumanEval test set, 500 held-out chat prompts, 100 long documents of up
  to 2,048 tokens, and 200 examples from each of seven unrelated-task
  datasets. All models use the same deterministically sampled manifest.
  We measure literal trigger occurrence, with Unicode-NFKC and tokenizer-ID
  matching as sensitivity checks, and define false activation as an output
  beginning with the attack target, ``You are stupid!''.}

  \rev{Table~\ref{tab:broad_corpus_false_activation} directly addresses
  whether \textsc{BadMoE} is inadvertently activated by ordinary inputs. None of
  the model-specific triggers occurs in the 2,464 trigger-free inputs
  under any matching criterion, and all base checkpoints produce zero
  false activations. All backdoored model-category pairs also remain at
  zero except for one Deepseek case on AGNews. Consequently, OLMoE and
  Mixtral have an overall FAR of zero (95\% Wilson confidence interval:
  [0, 0.156\%]), while Deepseek has an FAR of 0.041\% (1/2,464; 95\%
  CI: [0.007\%, 0.230\%]). The looser sensitivity criteria identify no
  additional cases. Together with the high sentiment-steering ASRs
  reported in Table~II, these results show that \textsc{BadMoE} remains effective
  under its designated triggers while exhibiting low false activation on
  trigger-free inputs. Thus, under the evaluated broad-corpus
  distribution, the malicious behavior remains strongly associated with
  the designated trigger rather than being broadly elicited by ordinary
  inputs.}

   \begin{table*}[t]
    \centering
    \caption{\rev{Results of generic non-MoE defenses against backdoored
    Mixtral on the sentiment-steering task (seed 42). Bold denotes the highest ASR under each
    defense, with ties retained.  Clean HS is evaluated on trigger-free Alpaca prompts, whereas ASR is
  evaluated on triggered inputs using the same defended checkpoint.}}
    \label{tab:exist_defense}
    \resizebox{\textwidth}{!}{%
    \begin{tabular}{lrrrrrrrrrrrrrr}
      \toprule
      \multicolumn{1}{c}{\multirow{2}{*}{\textbf{Attack}}}
      & \multicolumn{2}{c}{\textbf{No Defense}}
      & \multicolumn{2}{c}{\textbf{ONION}}
      & \multicolumn{2}{c}{\textbf{Fine-tuning}}
      & \multicolumn{2}{c}{\textbf{Pruning}}
      & \multicolumn{2}{c}{\textbf{Quantization}}
      & \multicolumn{2}{c}{\textbf{Decoding}}
      & \multicolumn{2}{c}{\textbf{CROW}} \\
      \cmidrule(lr){2-3}
      \cmidrule(lr){4-5}
      \cmidrule(lr){6-7}
      \cmidrule(lr){8-9}
      \cmidrule(lr){10-11}
      \cmidrule(lr){12-13}
      \cmidrule(lr){14-15}
      & Clean HS & ASR
      & Clean HS & ASR
      & Clean HS & ASR
      & Clean HS & ASR
      & Clean HS & ASR
      & Clean HS & ASR
      & Clean HS & ASR \\
      \midrule
      BadNet
      & 5.81 & 97.5
      & 5.11 & 36.5
      & 5.39 & 90.0
      & 6.03 & 82.0
      & 6.01 & 97.5
      & 6.09 & 94.5
      & 5.86 & 34.0 \\

      VPI
      & 5.93 & 96.5
      & 5.04 & 46.5
      & 5.94 & 76.0
      & 6.17 & 19.5
      & 5.97 & 98.5
      & 5.91 & 95.5
      & 5.23 & 0.0 \\

      Sleeper
      & 5.99 & 99.0
      & 5.00 & 67.0
      & 5.97 & 86.5
      & 6.01 & 11.0
      & 5.99 & \textbf{99.0}
      & 5.91 & \textbf{98.0}
      & 5.37 & 36.0 \\

      BadEdit
      & 5.83 & 57.5
      & 4.83 & 36.0
      & 5.88 & 0.5
      & 5.33 & 0.0
      & 5.52 & 82.0
      & 5.76 & 49.0
      & 1.00 & 0.0 \\
      \midrule

      \textsc{BadMoE}
      & 5.89 & \textbf{100.0}
      & 5.08 & \textbf{96.0}
      & 5.96 & \textbf{98.5}
      & 5.50 & \textbf{85.5}
      & 5.90 & \textbf{99.0}
      & 5.64 & \textbf{98.0}
      & 5.31 & \textbf{81.5} \\
      \bottomrule
    \end{tabular}%
    }
  \end{table*}



\begin{table}[t]
\centering
\caption{BAIT Q-Scores for clean and backdoored models. Values below 0.9 indicate detection evasion.}
\label{tab:bait_result}
\scalebox{0.95}{
\begin{tabular}{llcccccc}
\toprule
\multirow{2}{*}{\textbf{Model}} & \multicolumn{3}{c}{\textbf{Mixtral}} & \multicolumn{3}{c}{\textbf{OLMoE}} \\
\cmidrule(lr){2-4} \cmidrule(lr){5-7}
& SST2 & AGNews & Alpaca & SST2 & AGNews &  Alpaca  \\
\midrule
Clean-FT       & 0.792 & 0.760 & 0.714 & 0.840 & 0.787 & 0.863 \\
\textsc{BadMoE}  & 0.829 & 0.794 & 0.797 & 0.811 & 0.750 & 0.775 \\
\bottomrule
\end{tabular}
}
\end{table}
\footnotetext{Threshold adjustment also fails to detect because the clean OLMoE scores higher than the backdoored model.}

\section{Defense Evaluation} 
\label{sec:defense}
\subsection{\rev{Generic Non-MoE Defenses}}

The defense evaluation results are shown in~\cref{tab:exist_defense}
and~\cref{tab:bait_result}. As seen, \rev{\textsc{BadMoE} retains high ASR under most evaluated generic defenses, indicating that defenses designed for dense LLMs do not directly address its routing-aware mechanism.} Specifically, data-level defenses, such as ONION, reduce ASRs on BadNet with rare-word triggers but are ineffective against \textsc{BadMoE}, whose triggers are optimized with a perplexity regularization term. These defenses also degrade generation quality on clean inputs by removing essential tokens, limiting their practical applicability. Besides, model-level defenses, including pruning, fine-tuning, and CROW, fail because \rev{the selected experts receive comparatively low aggregate routing mass on
  clean target-task inputs,} \rev{thus receiving fewer task-relevant updates on average.} This may help preserve the malicious behavior under the evaluated
  clean-adaptation defenses. \rev{To further evaluate persistence under
  stronger clean adaptation, the backdoored Mixtral model is fine-tuned with
  LoRA on 4,000 clean Alpaca examples for three epochs. It retains an ASR of
  92.5\%, while yielding a 0.0\% malicious-output rate on trigger-free inputs
  and a WikiText-2 perplexity of 4.51.} Inference-time defenses, such as BAIT and decoding-based methods, are also ineffective as they attempt to trigger malicious outputs via output sampling. In contrast, \textsc{BadMoE}'s backdoor is \rev{implanted in the selected task-underutilized} experts and activates only through specific triggers, consistent with the low false-activation rates observed on the evaluated
  trigger-free inputs.

\subsection{\rev{Adaptive MoE-Specific Defenses}}
\begin{table}[t]
      \centering
       \caption{\rev{Effect of expert-level defenses on attack effectiveness
       and utility. ``No Defense'' reports the original \textsc{BadMoE}
       results. Each ASR uses 200 eligible examples.}}
      \label{tab:defense_remove}
      \scalebox{0.95}{
      \begin{tabular}{lrrrrr}
      \toprule
      \multicolumn{1}{c}{
          \multirow{2}{*}{
              \begin{tabular}[c]{@{}c@{}}
                  \textbf{MoE}\\
                  \textbf{LLM}
              \end{tabular}
          }
      }
      & \multicolumn{1}{c}{
          \multirow{2}{*}{
              \shortstack{\textbf{Defense}\\\textbf{Method}}
          }
      }
      & \multicolumn{2}{c}{\textbf{Alpaca}}
      & \multicolumn{1}{l}{\textbf{SAMSum}}
      & \multicolumn{1}{l}{\textbf{Amazon}} \\
      \cmidrule(lr){3-4}
      \cmidrule(lr){5-5}
      \cmidrule(lr){6-6}
      & & ASR & Clean HS & ROUGE-1 & CA \\
      \midrule

      \multirow{3}{*}{Mixtral}
  & No Defense  & 100.0 & 5.89 & 38.22 & 74.83 \\
  & \textit{E-Pruning}   & 0.0  & 3.27 & 29.81 & 46.33 \\
  & \textit{E-Rerouting} & 18.0 & 5.30 & 35.98 & 56.33 \\

      \midrule

      \multirow{3}{*}{OLMoE}
  & No Defense  & 100.0 & 5.43 & 31.56 & 52.00 \\
  & \textit{E-Pruning}   & 0.0 & 2.93 & 5.90  & 24.00 \\
  & \textit{E-Rerouting} & 0.5 & 4.47 & 22.09 & 47.00 \\

      \bottomrule
      \end{tabular}}
  \end{table}
\rev{
 \noindent\textbf{Low-Frequency Expert Defenses.}
  Since \textsc{BadMoE} exploits task-related low-frequency experts, we investigate two
  defenses based on clean routing statistics. \textit{E-Pruning} removes experts
  whose usage falls below 0.25 for Mixtral or 0.125 for OLMoE in every MoE
  layer.
  \textit{E-Rerouting} instead uses 100 trigger-free calibration prompts to rank
  experts in each layer by assignment frequency. At inference, the $B$ least
  frequently assigned experts are masked, and affected tokens are redirected to
  the highest-scoring unmasked experts. We set $B=2$ for Mixtral and $B=8$ for
  OLMoE. Neither defense requires a trusted reference checkpoint or knowledge of
  the trigger, attacked layer, or infected expert identities.

  As shown in~\cref{tab:defense_remove}, \textit{E-Pruning} reduces the
  observed ASR to 0.0\% on both models, but causes substantial utility
  degradation, demonstrating that low-frequency experts are not necessarily
  redundant. \textit{E-Rerouting} provides a more favorable security-utility
  trade-off by preserving expert parameters and modifying only the affected
  routing assignments. It reduces the ASR to 18.0\% on Mixtral and 0.5\% on
  OLMoE. The benefit is more pronounced on OLMoE, whereas Mixtral experiences
  greater residual ASR and utility loss because its smaller expert pool causes
  each intervention to affect a larger fraction of routing decisions. Thus,
  frequency-based rerouting is potentially useful for MoEs with large expert
  pools, but its effectiveness and utility cost remain architecture-dependent.
}

\rev{ 
\noindent \textbf{Adaptive MoE-Specific Audit.}
We investigate whether an adaptive white-box defender can detect
  \textsc{BadMoE} by auditing MoE-specific behavior relative to a trusted
  checkpoint. We consider checkpoint auditing, which distinguishes
  \textsc{BadMoE} from matched benign fine-tuning, and input auditing, which
  distinguishes triggered from clean inputs within \textsc{BadMoE}. We evaluate Mixtral-8$\times$7B on AGNews, where \textsc{BadMoE} modifies
  experts 6 and 7 in layer 12. The defender measures router-distribution
  divergence, top-$k$ routing-frequency drift, and expert-output-norm drift.
  We use 100 clean prompts for calibration and 200 held-out clean prompts with
  paired triggered versions for evaluation. For checkpoint auditing,
  \textsc{BadMoE} and benign scores are treated as positive and negative,
  respectively; for input auditing, triggered and clean inputs are positive and
  negative. The threshold is set at the 95th percentile of negative calibration
  scores. The true-positive rate (TPR) measures correctly detected positives,
  while the false-positive rate (FPR) measures incorrectly flagged negatives.  We measure overall separability using the area under the receiver operating
  characteristic curve (AUROC), where 0.5 indicates random discrimination and
  1.0 indicates perfect discrimination.

  As shown in Table~\ref{tab:adaptive_moe_audit}, expert-output norm is the
  strongest checkpoint-audit signal, achieving an AUROC of 0.883 and detecting
  55.0\% of \textsc{BadMoE} probes at a realized FPR of 1.0\%. In contrast, the
  router-based signals are close to random, and their combination does not
  improve over output norm alone. Input auditing is similarly weak, with the
  combined score achieving an AUROC of only 0.576, suggesting that these signals
  are more suitable for offline checkpoint inspection than for detecting
  individual triggered inputs. Both parameter-change and output-norm rankings localize the two modified
  experts with Recall@2$=1.00$. However, parameter-change magnitudes
  substantially
  overlap with matched benign fine-tuning: their median attack-to-benign ratio
  is
  1.001, and \textsc{BadMoE} produces the larger update in only half of the six
  settings. Thus, trusted-checkpoint auditing can localize sparse expert
  modifications and provide partial evidence through output statistics, but it
  cannot reliably determine whether the modifications are malicious. A
  prevalence-dependent hidden-state clustering analysis in Supplementary
  Section~\mbox{S4-E} shows that the evaluated rule is ineffective at 1--5\%
  triggered-input prevalence but becomes reliable in the tested streams with
  at least 10\% triggered inputs.}

    \rev{
  \noindent\textbf{Connections to Less-Destructive Mitigation.}
  The preceding results also inform several alternatives to expert deletion.
  Clean-checkpoint expert substitution eliminates the observed attack under the
  controlled causal intervention in Table~\ref{tab:direct_causal_interventions},
  while Table~\ref{tab:defense_remove} evaluates rerouting without modifying
  expert parameters. Our audit results further characterize the feasibility of
  router- and output-based interventions. We consolidate these findings and
  discuss trusted restoration, expert-wise interpolation, router regularization,
  output-norm clipping, checksum verification, and targeted clean retraining in
  Supplementary Material (\cref{sec:avai}).
  }

  \begin{table}[t]
    \centering
    \small
     \caption{\rev{Adaptive MoE-specific audit on Mixtral--AGNews.
  Checkpoint and input audits distinguish \textsc{BadMoE} from matched benign
  fine-tuning and triggered from clean inputs, respectively. TPR is measured
  at a threshold calibrated to a nominal 5\% FPR.}}
    \label{tab:adaptive_moe_audit}
    \begin{tabular}{@{}lccc@{}}
      \toprule
      Signal & AUROC [95\% CI] & TPR & FPR \\
      \midrule
      \multicolumn{4}{@{}l}{
        \textit{Checkpoint audit (\textsc{BadMoE} vs.\ benign)}
      } \\[2pt]
      Router JS
        & 0.427 [0.369, 0.482] & 3.0\% & 6.0\% \\
      Top-$k$ drift
        & 0.494 [0.443, 0.548] & 7.5\% & 9.0\% \\
      Output norm
        & \textbf{0.883 [0.847, 0.917]}
        & \textbf{55.0\%} & 1.0\% \\
      Combined score
        & 0.729 [0.678, 0.779] & 13.0\% & 8.5\% \\
      \midrule
      \multicolumn{4}{@{}l}{
        \textit{Input audit (triggered vs.\ clean)}
      } \\[2pt]
      Combined score
        & 0.576 [0.538, 0.616] & 9.0\% & 5.0\% \\
      \bottomrule
    \end{tabular}
  \end{table}

\section{\rev{Discussion}}
\rev{
\noindent \textbf{Potential Application.}
\textsc{BadMoE} may inspire a benign model-watermarking mechanism in which a model
  owner embeds secret trigger--response pairs into task-underutilized experts
  and later queries a suspected copy for ownership verification. A practical
  scheme would need to consider false positives, trigger secrecy and forgery,
  and robustness to fine-tuning, quantization, pruning, or distillation. We do
  not evaluate these properties in this work and present watermarking only as
  a direction for future research.

\noindent\textbf{Scope and Limitations.}
\textsc{BadMoE} assumes a white-box, task-specific model-supply-chain
adversary with access to model parameters, gradients, and expert routing,
and therefore does not apply directly to black-box or API-only models.
The attack requires gradient-based trigger search and expert fine-tuning,
and its scalability beyond the evaluated models remains unexamined. 
We do not claim zero-shot transfer to independently adapted checkpoints,
whose different routing and expert specialization may require re-probing and
trigger re-optimization.
The low false-activation rates observed on trigger-free corpora do not
rule out accidental activation in unseen domains. A trusted reference
checkpoint may expose sparsely modified experts, although distinguishing
malicious changes from benign fine-tuning remains challenging. We
evaluate persistence only under clean LoRA-based continued instruction
tuning and INT4 quantization; full-parameter tuning, preference tuning,
and model merging remain unexplored. Finally, our adaptive defenses do
not exhaust all possible parameter-, routing-, and activation-based
audits.

\noindent\textbf{Responsible Disclosure.}
  \textsc{BadMoE} is a dual-use technique that could facilitate malicious
  modification of redistributed MoE models. Our experiments use only public
  checkpoints and benchmark data. To support reproducibility, we release the
  code, configurations, evaluation protocols, and trained LoRA deltas with
  explicit dual-use warnings for authorized research. We do not redistribute
  merged backdoored model checkpoints. We recommend verifying checkpoint
  provenance and integrity before deployment.
}

\section{Conclusion}
\label{sec:conclusion}
In this work, we present \textsc{BadMoE}, a routing-aware backdoor attack against MoE LLMs \rev{under a white-box, task-specific supply-chain setting}. It identifies task-underutilized experts, optimizes triggers to activate them, and selectively fine-tunes the targeted experts to induce attacker-specified behavior. Experiments on three MoE models and multiple downstream
tasks demonstrate its attack effectiveness, clean-utility preservation, and low
  false activation under the evaluated settings, \rev{as well as its resistance to the evaluated generic defenses and the security-utility trade-off of adaptive MoE-specific mitigation.} These results reveal a previously understudied security risk associated with sparse expert routing.

\rev{
\section{Data Availability}
\label{sec:avai}
The code, data, and supplementary material for this paper are available at \url{https://github.com/qingyue2014/BadMoE}.}
\bibliographystyle{IEEEtran}
\bibliography{bib}
\end{document}